\documentclass[twocolumn,pra,showpacs,superscriptaddress,nofootinbib,notitlepage,amssymb,preprintnumbers]{revtex4-1}
\usepackage{amsmath,amssymb,amsthm}
\usepackage{braket}
\usepackage{graphicx}
\usepackage[pdfstartview=FitH]{hyperref}
\usepackage[export]{adjustbox}
\usepackage{empheq}
\usepackage{subfigure}

\hypersetup{
    colorlinks=true,           linkcolor=blue,             citecolor=magenta,            filecolor=magenta,          urlcolor=cyan,               runcolor=cyan
}

\newtheorem*{theorem*}{Theorem}

\newtheorem{mylemma}{Lemma}

\newcommand{\bes} {\begin{subequations}}
\newcommand{\ees} {\end{subequations}}
\newcommand{\beq} {\begin{equation}}
\newcommand{\eeq} {\end{equation}}

\newcommand{\Om}{\Omega}
\newcommand{\dt}{\delta t}
\newcommand{\Texp}{\text {Texp}}
\newcommand\norm[1]{\left\lVert#1\right\rVert}

\begin{document}

\title{Suppression of effective noise in Hamiltonian simulations}

\begin{abstract}
Simulating high-weight Hamiltonians
 can convert local noise on the original Hamiltonian into undesirable nonlocal noise on the simulated Hamiltonian. 
Here we show how starting from two-local Hamiltonian in the presence of non-Markovian noise, a desired computation can be simulated as well as protected using fast pulses, while maintaining an energy gap against the errors created in the process.
\end{abstract}

\author{Milad Marvian}
\affiliation{Department of Electrical Engineering, University of Southern California, Los Angeles, California 90089, USA}
\affiliation{Center for Quantum Information Science \&
Technology, University of Southern California, Los Angeles, California 90089, USA}

\author{Todd Brun}
\affiliation{Department of Electrical Engineering, University of Southern California, Los Angeles, California 90089, USA}
\affiliation{Center for Quantum Information Science \&
Technology, University of Southern California, Los Angeles, California 90089, USA}
\affiliation{Department of Physics and Astronomy, University of Southern California, Los Angeles, California 90089, USA}

\author{Daniel A. Lidar}
\affiliation{Department of Electrical Engineering, University of Southern California, Los Angeles, California 90089, USA}
\affiliation{Center for Quantum Information Science \&
Technology, University of Southern California, Los Angeles, California 90089, USA}
\affiliation{Department of Physics and Astronomy, University of Southern California, Los Angeles, California 90089, USA}
\affiliation{Department of Chemistry, University of Southern California, Los Angeles, California 90089, USA}

\maketitle

\section{Introduction}
The theory of quantum fault tolerance ensures that quantum computers can operate reliably in the presence of decoherence and noise \cite{shor1996fault,lidar2013quantum}. The quantum accuracy threshold theorem, in various incarnations \cite{aharonov1997fault,kitaev1997quantum,Knill:98,Gottesman:99,Knill:05,aliferis_quantum_2006,reichardt_fault-tolerance_2005,aharonov_fault-tolerant_2006,ng:032318,ng_combining_2011,Gottesman:2013ud}, guarantees that arbitrary long, reliable quantum computation is achievable if the error rate is below a threshold. Although in principle the existence of this threshold means that scalable quantum computation is possible, in practice the value of this threshold and the required overhead are very important, as achieving them in experiments remains extremely challenging.

One approach to reduce these requirements is to use active quantum error correction in combination with other methods that provide additional robustness against instability or noise. 
Among such methods are holonomic quantum computation \cite{zanardi1999holonomic} and topological quantum computation \cite{kitaev2003fault}.
In \cite{zheng_fault2015,cesare_adiabatic_2015} schemes have been proposed based on encoding the information in the ground subspace of a Hamiltonian with a constant energy gap and topological properties, such as the surface code Hamiltonian, and performing the computation by adiabatically deforming the Hamiltonian. The energy gap of the Hamiltonian suppresses the thermal excitations induced by the environment. Compatible active error correction is performed frequently enough to prevent logical errors. As the energy gap protection from a Hamiltonian 
stabilizer code on a two-dimensional lattice cannot increase with system size \cite{bravyi_no-go_2009}, to sustain arbitrary long universal quantum computation on such a lattice with stabilizer codes requires active error correction.

Although this construction is appealing, the surface code Hamiltonian and also the time-dependent Hamiltonian implementing the deformation consist of up to $4$-local interactions \cite{zheng_fault2015}. Implementing such interactions experimentally is difficult (though not impossible and various proposals exist \cite{Mizel:2004ve,Chancellor:2016rz,Puri:2016aa}). One way to circumvent this is to simulate these Hamiltonians using other available resources. 
But the noise of the original system may be dramatically transformed by the simulation procedure. It is important to investigate how the simulation process converts realistic noise on the original resources to new effective noise on the simulated Hamiltonian. For example, local noise on the original Hamiltonian can be converted into nonlocal noise on the simulated Hamiltonian, and this could reduce the effectiveness of the simulated Hamiltonian in suppressing errors. This is a general concern for various simulation methods, including stroboscopic methods 
(see, e.g., \cite{lidar2001decoherence,dodd_universal_2002}) and perturbative gadgets \cite{Jordan:08,Cao:2014,Subasi:2016fk}.

The question of the effect of a Markovian noise on simulation has been partially studied \cite{dur_quantum_2008}, but despite its importance  \cite{cirac_goals_2012}, 
only a limited amount of attention has been devoted to  the effects of general noise on the simulation of high-weight Hamiltonians. 
In this work we show how---starting from an 
entangling Hamiltonian in the presence of a general local non-Markovian environment---one can generate the desired nonlocal Hamiltonian by the application of one-local unitary operators.
By combining 
simulation with schemes for dynamical decoupling (DD) \cite{viola1998dynamical,viola_dynamical_2000,lidar2013quantum}, we construct a sequence of pulses that suppresses errors while simulating the desired interactions. The relationship between the strength of the simulated Hamiltonian and the strength of effective noise is investigated. We also consider how far the transformed errors spread, based on the locality (both geometric and algebraic) of the bath Hamiltonian and the system-bath interaction.

We illustrate our construction with the Hamiltonians used in Ref.~\cite{zheng_fault2015} for fault-tolerant quantum computation in the surface code, but it can easily be modified to simulate other, similar Hamiltonians, including the Hamiltonians used for protection of adiabatic quantum computation in Ref.~\cite{jordan_error-correcting_2006}. In the latter, a stabilizer code that can detect the effect of the environment on the system is chosen, the system Hamiltonian is then encoded using the logical operators, while a penalty Hamiltonian is added to break the degeneracy. Again, the energy gap of the constructed Hamiltonian suppresses the rate of excitation out of the code space \cite{Marvian:2017aa}. This scheme also requires $4$-body interactions.

\section{Summary of main results and organization of the paper}
Our main contributions in this work can be summarized as follows:

\begin{itemize}
  \item We provide a method to simulate Hamiltonians with high-weight interactions on a grid, such as surface code Hamiltonians, in a number of steps that is independent of the size of the grid. We demonstrate this with detailed construction steps for surface code Hamiltonians, using two different stroboscopic simulation techniques.
  
  \item  We provide a method to design new pulse sequences that can simulate a desired Hamiltonian while pushing errors (caused by the presence of the bath) to higher orders, by combining DD and Hamiltonian simulation techniques. 
   To do so, we use two application of the symmetrization procedure \cite{viola_dynamical_2000,zanardi_symmetrizing_1999}. The first application shows how to protect a Hamiltonian while averaging out all the errors. The second application reduces the number of pulses needed to average out local errors.
   
  \item For local, but otherwise general, non-Markovian noise on the original Hamiltonian we 
show how the effective error on the simulated Hamiltonian becomes non-local. We demonstrate that this effective noise is mostly suppressed during the simulation of a surface code Hamiltonian because of an effective gap.
  
\end{itemize}

This paper is organized as follows. In Sec.~\ref{setup} we introduce the resources and the goal of the simulation. In Sec.~\ref{sec:noisefree}, we describe two different methods to simulate the desired high-weight Hamiltonian on a grid in the noise-free setting. We also discuss how to simulate a time-dependent Hamiltonian that, with the help of active error correction, can perform universal quantum computation fault-tolerantly. In Sec.~\ref{sec:noise}, we discuss the effects of general non-Markovian noise on the simulation. We then show how using DD techniques in the simulation procedure can reduce the strength of the effective noise. We also discuss how the locality of the bath and system-bath interaction Hamiltonian changes the spread of the noise and also the resources required for a successful simulation. 
In Sec.~\ref{sec:errorsup}, we explicitly show that during the simulation an effective gap against the strongest errors created in the process is maintained. Section \ref{sec:conclusion} is devoted to conclusion and discussion. Additional technical details are provided in the Appendix.

\section{Setup} \label{setup}
\begin{figure}
\centering
\includegraphics[width=0.8\linewidth]{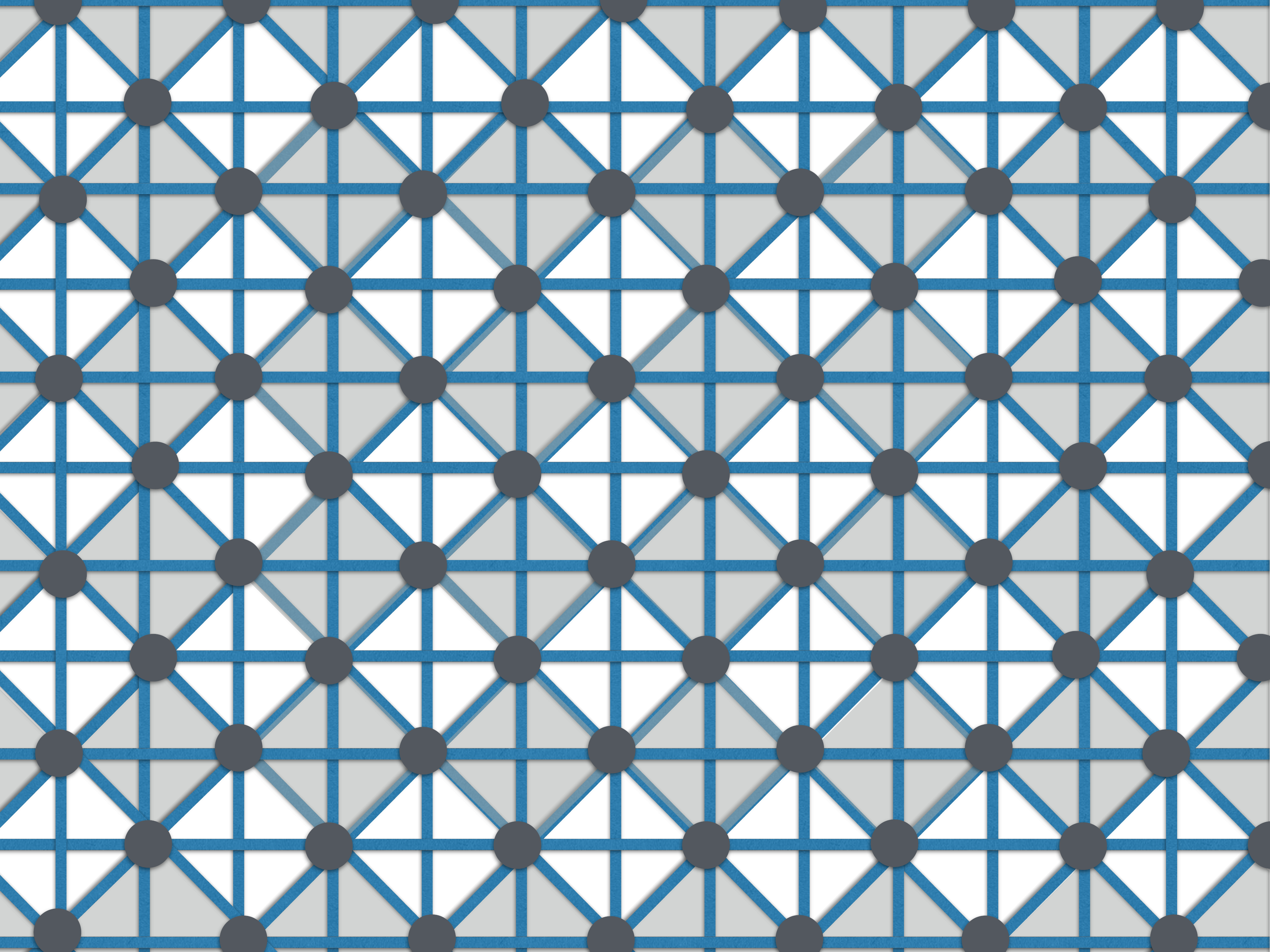}
\caption{Interacting qubits on a grid. Blue lines represent $XX$ interactions between qubits (circles). There is an interaction between nearest neighbor and next-nearest neighbor qubits. For the toric Hamiltonian, the goal is to simulate $X^{\otimes 4}$ on the shaded areas and $Z^{\otimes 4}$ on the light areas.}
\label{FigGrid}
\end{figure}

Assume $N$ qubits are placed on a two-dimensional square lattice as depicted in Fig.~\ref{FigGrid}. Let $d(i,j)$ denote the Euclidean distance between qubits $i$ and $j$; we choose units such that $d(i,j)=1$ if qubits $i$ and $j$ are nearest neighbors. A two-body entangling Hamiltonian acts on nearest neighbor and next-nearest neighbor qubits (with $d(i,j)=\sqrt{2}$):
\begin{eqnarray} \label{eq:H_X}
H_X=\sum_{\langle\langle i,j\rangle\rangle}{X_i X_j}\ ,
\end{eqnarray}
where $X_i$ denotes the Pauli $\sigma^x$ matrix acting on qubit $i$. (An example of an alternative connectivity is provided in Appendix ~\ref{App:alternativeconnectivity}.) 

Let us start with the toric code Hamiltonian; later we will convert it to the surface code Hamiltonian with cuts. In this case the Hamiltonian is:
\begin{eqnarray}
\label{eq:H_p}
H_p= \sum_{v}{A_v} + \sum_{p}{B_p},
\end{eqnarray}
with each vertex operator $A_v$ corresponding to an $X^{\otimes 4}$ term acting on the spins connected to the vertex $v$ and each plaquette operator $B_p$ corresponding to a $Z^{\otimes 4}$ term acting on the spins in the plaquette $p$; see Fig.~\ref{FigGrid} ($Z$ denotes the Pauli $\sigma^z$ matrix). This Hamiltonian is $4$-local, and as the first step we show how to simulate it in the ideal case (no noise) using the $2$-local Hamiltonian $H_X$.

Having stroboscopic simulation in mind, for now we assume that we can apply instantaneous, ideal unitary pulses on each qubit. For simplicity we assume that the time interval between pulses is fixed (more general schemes can improve the results).

\section{Simulation: Ideal case} \label{sec:noisefree}
Assuming there is no noise, here we show how to use the commutator method to simulate the Hamiltonian $H_p$ of Eq.~\eqref{eq:H_p} using the Hamiltonian $H_X$ of Eq.~\eqref{eq:H_X}.  We set $\hbar\equiv 1$.

\subsection{Commutator method}
One way to build a $3$-local Hamiltonian using $2$-local Hamiltonians is to use the identity:
\begin{eqnarray} 
\label{eq:commutator2id}
\Omega_\tau(A,B) &\equiv& e^{iB\tau}e^{iA\tau}e^{-iB\tau}e^{-iA\tau} \nonumber \\
&=&e^{-i  \tau^2 (i[A,B])}+O(\tau^3)\ ,
\end{eqnarray}
valid for any pair of operators $A$ and $B$. This allows us to generate an effective $3$-local Hamiltonian $i\delta t[H_a,H_b]$ using appropriate $2$-local Hamiltonians $H_a$ and $H_b$:
\begin{eqnarray} 
\label{eq:commutator2}
\Omega_{\delta t}(H_a,H_b)=e^{-i  \delta t^2 (i[H_a,H_b])}+O(\delta t^3)\ .
\end{eqnarray}

Note that in oder to neglect the higher order terms, obviously $\|H_a\|\delta t$ and $\|H_b\|\delta t$ must be small. Also, the effective Hamiltonian contains $\delta t$, and so is much weaker than the Hamiltonians we started with. 

The same method can be used iteratively to increase the locality of the effective Hamiltonian. To generate a $4$-local Hamiltonian we notice that by negating $H_b$ or $H_a$, the term $[H_a,H_b]$ is also negated. So we repeat the procedure with $H_a$, $-H_b$, and a new Hamiltonian $H_c$:
\begin{eqnarray} 
\label{Eq:commutatorMethod}
U&=& e^{iH_c \delta t}\Om_{\dt}(H_a,-H_b) e^{-iH_c \delta t}\Om_{\dt}(H_a,H_b) \nonumber \\
&=&e^{-i \delta t^3 ([[H_a,H_b],H_c]+[[H_a,H_b],H_b]])}+O(\delta t^4)\ .
\end{eqnarray}
The term $\delta t^2[[H_a,H_b],H_c]$ is the desired $4$-local Hamiltonian. The extra term, $\delta t^2[[H_a,H_b],H_b]$, is the result of the third-order error from  Eq.~\eqref{eq:commutator2} which becomes relevant now:
\begin{eqnarray}
&&\Omega_{\delta t}(H_a,H_b) \nonumber \\
&&= e^{-i \delta t^2 (i[H_a,H_b]) -i\delta t ^3  [[H_a,H_b],H_a+H_b]}+O(\delta t^4)\ , 
\label{eq:2commutatormethod}
\end{eqnarray}
where the third-order error is not negated by the replacement $H_b  \mapsto - H_b$; the term $[[H_a,H_b],H_b]$ remains. Using the operator identity
\beq
\label{eq:conjugate-by-u}
u e^{A} u^\dagger = e^{uAu^\dagger} = 
\begin{cases}
e^{A} & [u,A]=0\\
e^{-A} & \{u,A\}=0
\end{cases}
\eeq
valid for unitary $u$ and arbitrary $A$, one can apply a pulse $u$, itself generated by a $2$-local Hamiltonian, that commutes with the desired Hamiltonian and anticommutes with the extra term to eliminate it:
\begin{eqnarray} \label{eq:removeextraterms}
UuUu^{\dagger}=e^{-2i \delta t^3  [[H_a,H_b],H_c]}+O(\delta t^4)\ .
\end{eqnarray}
This process takes time $20\delta t$. {This type of simulation is stroboscopic, so we obtain the desired effective Hamiltonian at a specific time (here at $t=20 \delta t$).}

In the above discussion we used three different Hamiltonians $H_{a,b,c}$. For the actual simulation, we will assume that $H_X$ is always on, and then apply pulses (not generated by $H_X$) to generate the effective Hamiltonians $H_{\{a,b,c\}}$. 

\subsection{Simulating one plaquette operator} \label{Sec:SimOneOperator}
\begin{figure}
\begin{center}
\includegraphics[height=0.2\textheight]{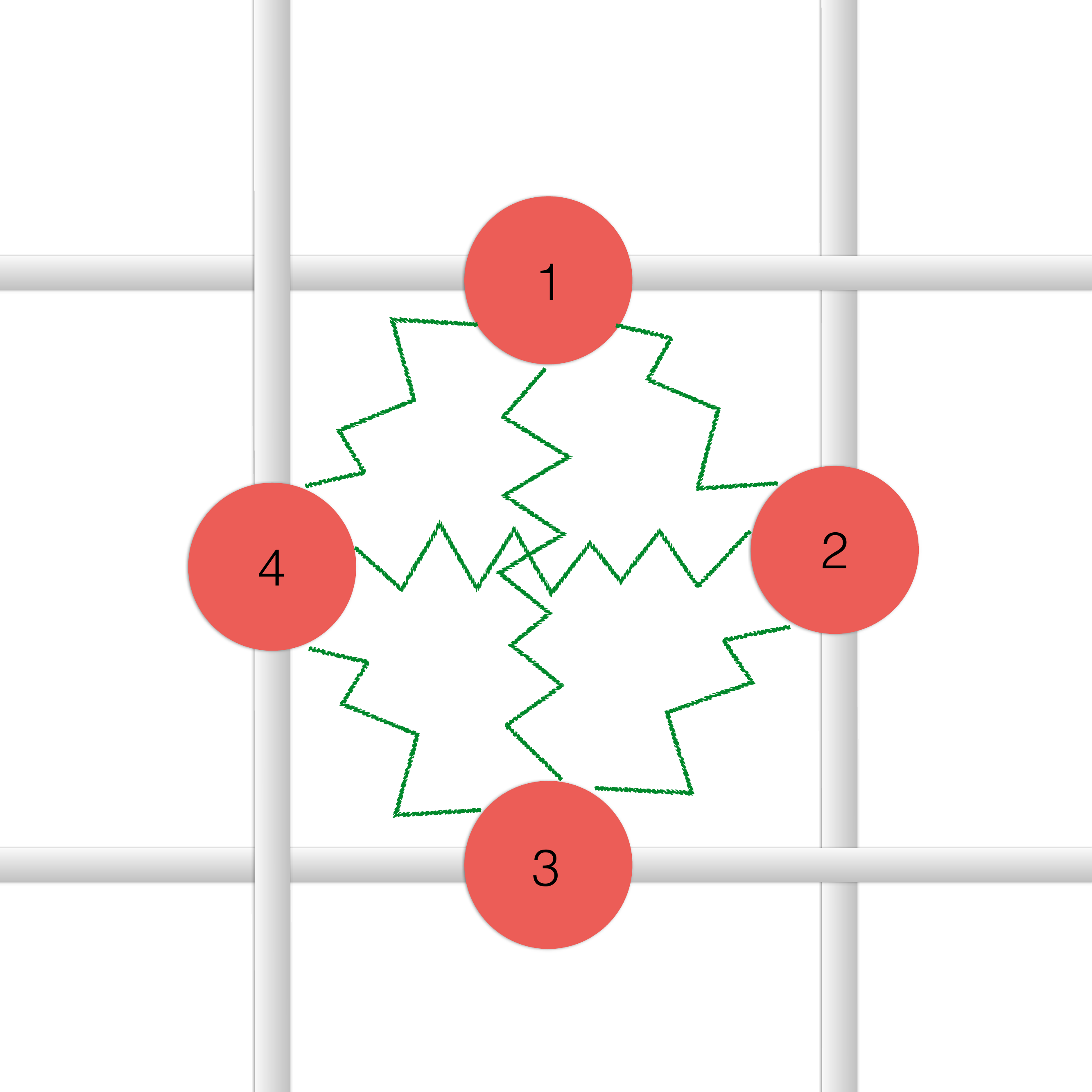}
\caption{Depiction of $H_x= \sum_{1\leq i < j \leq 4}{X_i X_j}$ for four qubits in a plaquette.}
\label{fig:OnePlaquette}
\end{center}
\end{figure}

Let us now show how to use the Hamiltonian 
\beq
H_x= \sum_{1\leq i < j \leq 4}{X_i X_j}
\eeq 
(not to be confused with $H_X$) to simulate the Hamiltonian $X_1 X_2 X_3 X_4$ on the qubits of a plaquette (see Fig.~\ref{fig:OnePlaquette}).

We will repeatedly use some basic identities, which are listed next for convenience, as they will be used throughout the remainder of this work:
\begin{eqnarray}
WXW &= Z, \qquad WYW =-Y,  \qquad WZW = X, \nonumber \\
SXS^{\dagger} &=Y, \qquad SYS^{\dagger} =-X, \qquad SZS^{\dagger}=Z,
\end{eqnarray}
where $W$ is the Hadamard gate and $S$ is the phase gate.

Let the system evolve with the Hamiltonian $H_x$ for a time $2\delta t$, and apply $W_1 W_2, u=Z_1 Z_2$ and $W_1 W_2 u$ pulses at times $t=0$, $\delta t$ and $2 \delta t$ respectively, where $W_i$ is a Hadamard pulse on the $i$-th qubit:
\begin{equation}
W_1 W_2(u e^{-i H_x \delta t} u) e^{-i H_x \delta t} W_1 W_2 =e^{-i 2(Z_1 Z_2+X_3 X_4) \delta t}	.
\end{equation}
The resulting Hamiltonian is $H_a=2(Z_1 Z_2+X_3 X_4)$.

We generate $H_b=2(Y_1 X_4+X_2 X_3)$ by first applying the inverse phase gate $S^{\dagger}_1$ to the first qubit, then applying $u=Z_1 Z_4$ at time $\delta t$, and finally $S_1u$ at time $2 \delta t$:
\begin{eqnarray}
S_1 (u e^{-i H_x \delta t} u) e^{-i H_x \delta t} S^{\dagger}_1 =e^{-i 2(Y_1 X_4+X_2 X_3) \delta t}\ .
\end{eqnarray}

We also generate $H_c=2(Y_2 X_3+X_1 X_4)$ by applying pulses $S^{\dagger}_2$, $u=Z_1 Z_4$, and $S_2u$ at times $0, \delta t$ and $2 \delta t$:
\begin{eqnarray} \label{eq:generateHc}
S_2 (u e^{-i H_x \delta t} u) e^{-i H_x \delta t} S^{\dagger}_2 =e^{-i 2(X_1 X_4+Y_2 X_3) \delta t}\ .
\end{eqnarray}

Generating any of  $-H_{a,b,c}$ is almost the same as generating $H_{a,b,c}$, but with a few additional pulses. For example, for $H_a$ we notice that:
\begin{equation}
(X_1 Z_3) e^{-i 2(Z_1 Z_2+X_3 X_4) \delta t}  (X_1 Z_3)=e^{i 2(Z_1 Z_2+X_3 X_4) \delta t}\ .
\end{equation}

In the same way one can generate $-H_b$ by conjugating $H_b$ with $X_1 Z_2$ pulses, and $-H_c$ by conjugating $H_c$ with $Z_1 Z_2$ pulses.

Having all the ingredients needed for Eq.~\eqref{Eq:commutatorMethod}, the effective Hamiltonian at order $\delta t^3$ contains the desired 4-body Hamiltonian with some extra terms that can be removed with only one extra step. To see this, note that:
\bes
\begin{align}
{[[H_a,H_b],H_c]}&= 32( X_1 X_2 X_3 X_4+ Y_1 Y_2 X_3 X_4)\ ,\\
{[[H_a,H_b],H_b]]}&= 64 (Z_1 Z_2 + X_1 Y_2 X_3 X_4)\ .
\end{align}
\ees
One then follows Eq.~\eqref{eq:removeextraterms} with an extra pulse $u=Z_1 Y_2$, which commutes with the desired term ($X_1 X_2 X_3 X_4$) and anticommutes with all the other terms. The effective Hamiltonian becomes $64 \delta t^2 X_1 X_2 X_3 X_4$, as desired.

Therefore, in $N_{X}=40$ steps we can convert the always-on Hamiltonian $H_x$ (after a time $N_X\delta t$) to $64 \delta t^2 X_1X_2X_3X_4$, with errors of order $\delta t^4$, which is the same as saying that the original Hamiltonian is stroboscopically converted to a $4$-local Hamiltonian at the price of making it $64 \delta t^2 /N_{X}$ times weaker.

\begin{figure*}[t]
\subfigure[\ ]{\includegraphics[width = 0.24\linewidth]{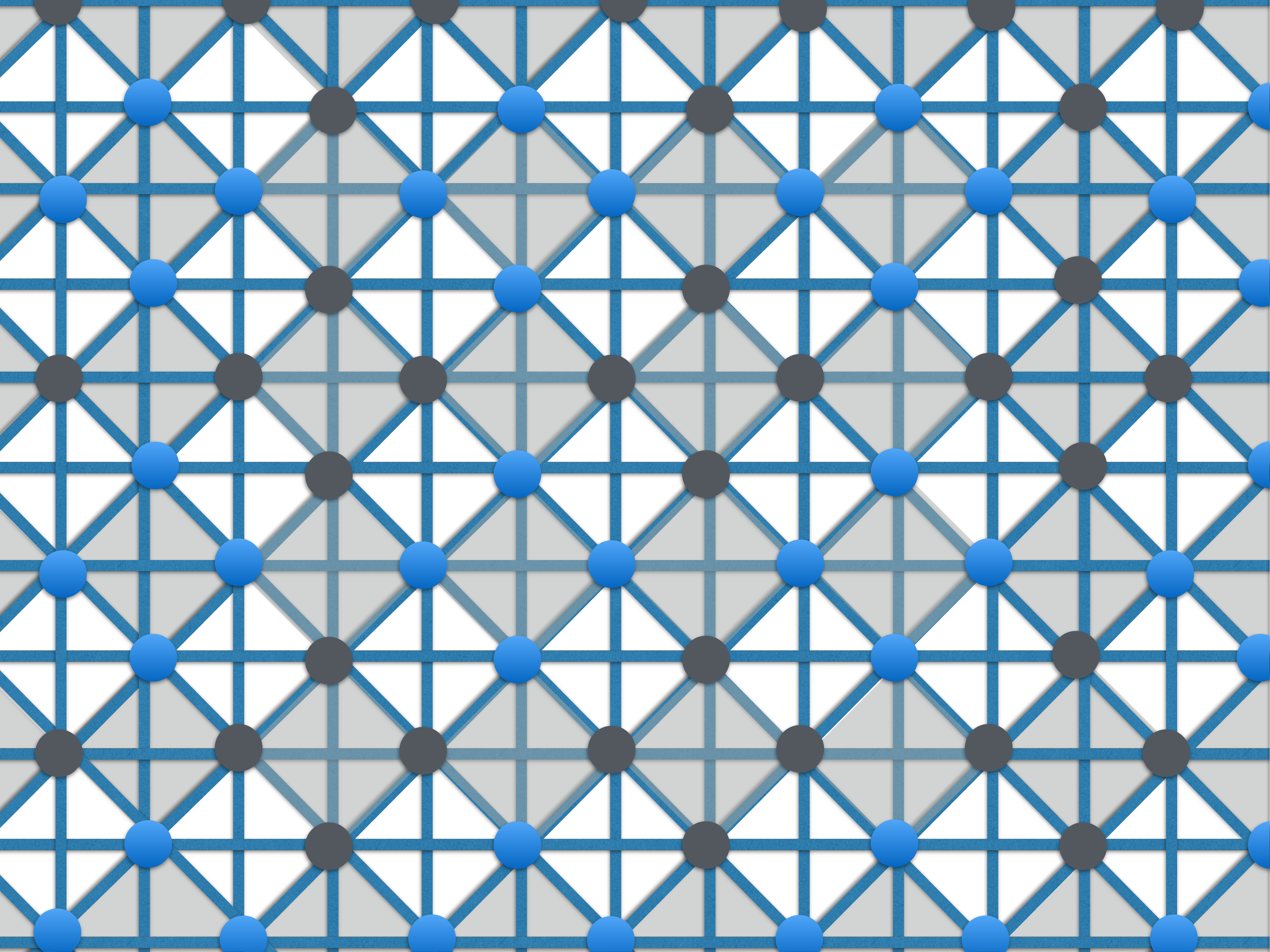}\label{FigGrid_Htemp_Zpulses}}
\subfigure[\ ]{\includegraphics[width = 0.24\linewidth]{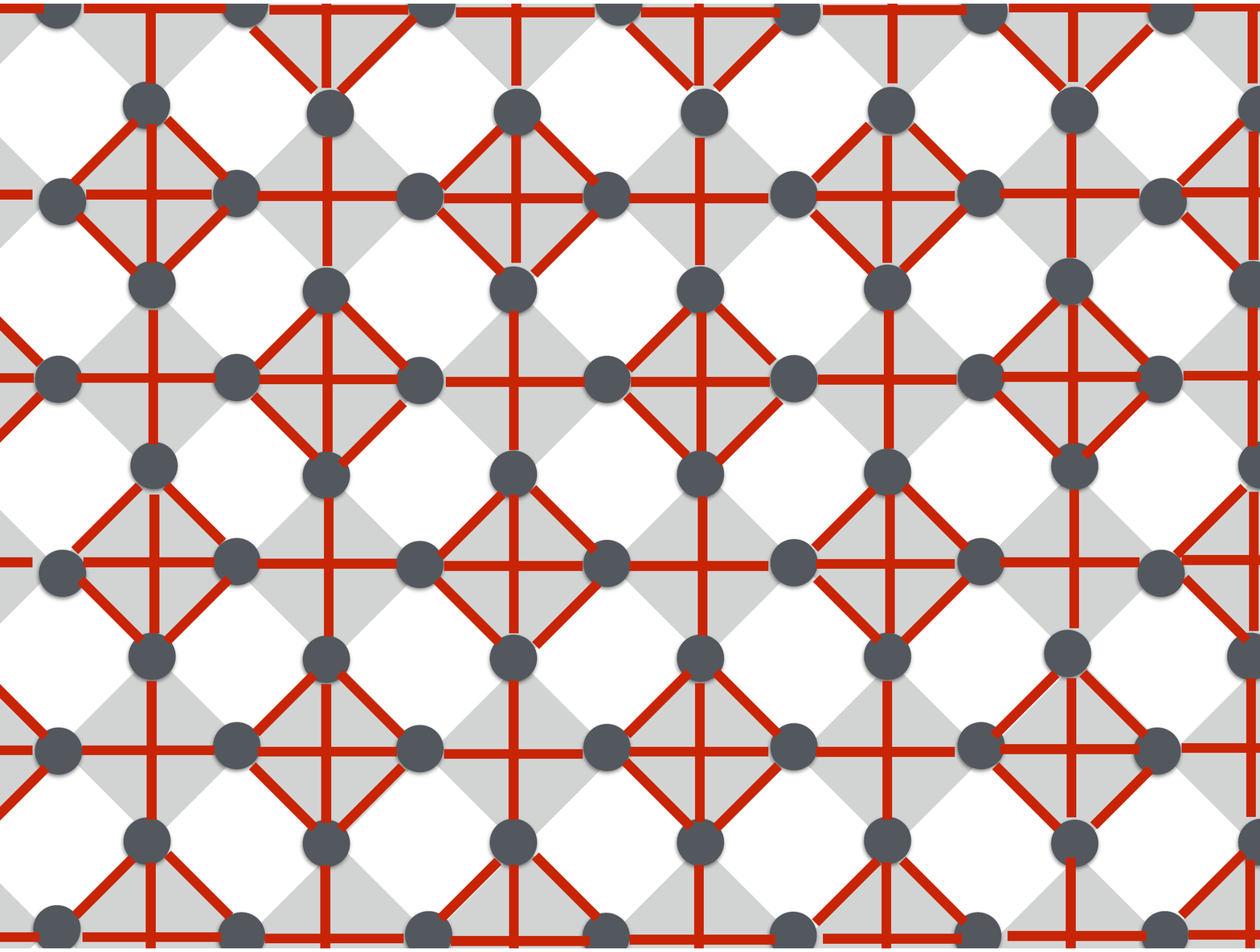}\label{FigGrid_Htemp}}
\subfigure[\ ]{\includegraphics[width = 0.24\linewidth]{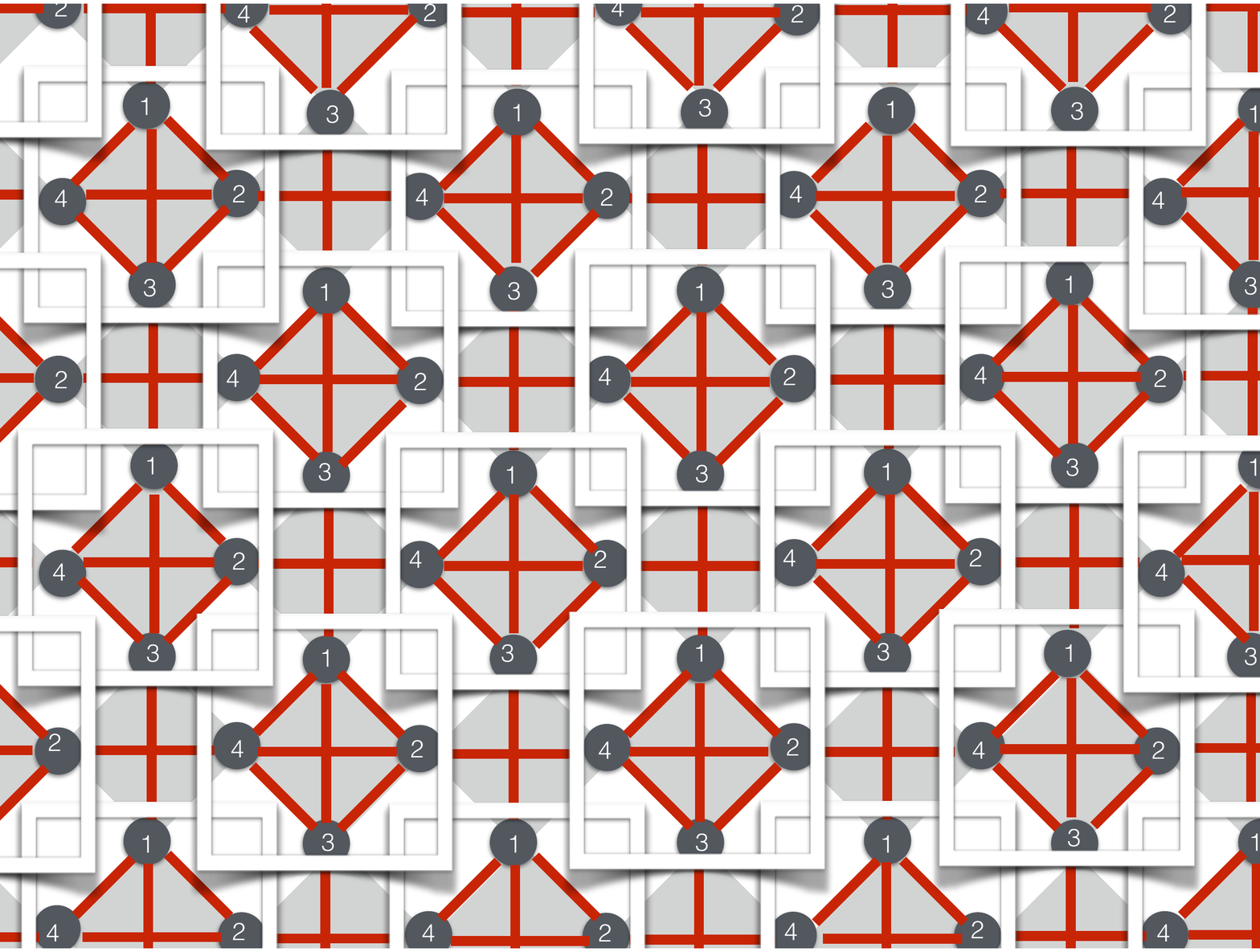}\label{FigGrid_Htemp_Labeld}}
\subfigure[\ ]{\includegraphics[width = 0.24\linewidth]{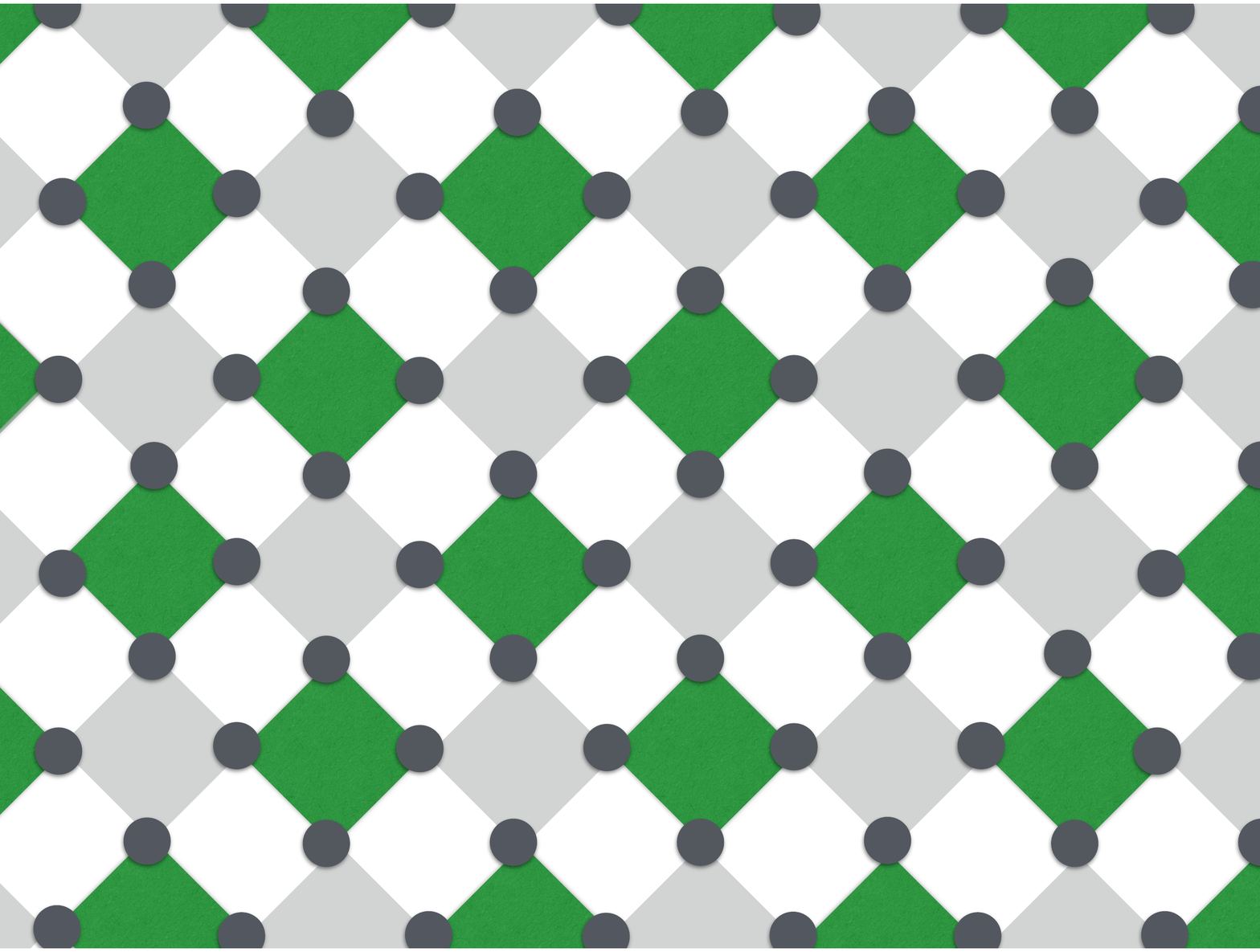}\label{FigGrid_4Body}}
\caption{Generating plaquette operators. (a) The location of $Z$ pulses is shown in blue. (b) The remaining $XX$ interactions after (a) are shown in red. Their sum defines $H_{\mathrm{temp}}$. (c) Labelling of qubits according to Fig.~\ref{fig:OnePlaquette}. (d) The result after applying pulses as is Sec.~\ref{Sec:SimOneOperator} to (c). This achieves the first goal of generating plaquette operators in the green areas. 
}
\label{fig:test}
\end{figure*}

\subsection{Simulating all the plaquette operators on the grid} 
\label{sec:SimulatingAllPlaquettes}

We assume that simultaneous pulses on all the qubits of the grid can be applied in parallel. Using this we can simulate all the plaquette operators on the grid. To do so, we start with the Hamiltonian $H_X$ and repeat the steps provided for one plaquette operator, while ensuring that no unwanted terms are created in the process.

\begin{figure*}[t]
\centering
\hspace*{\fill}
\subfigure[\ ]{\includegraphics[width = 0.24\linewidth]{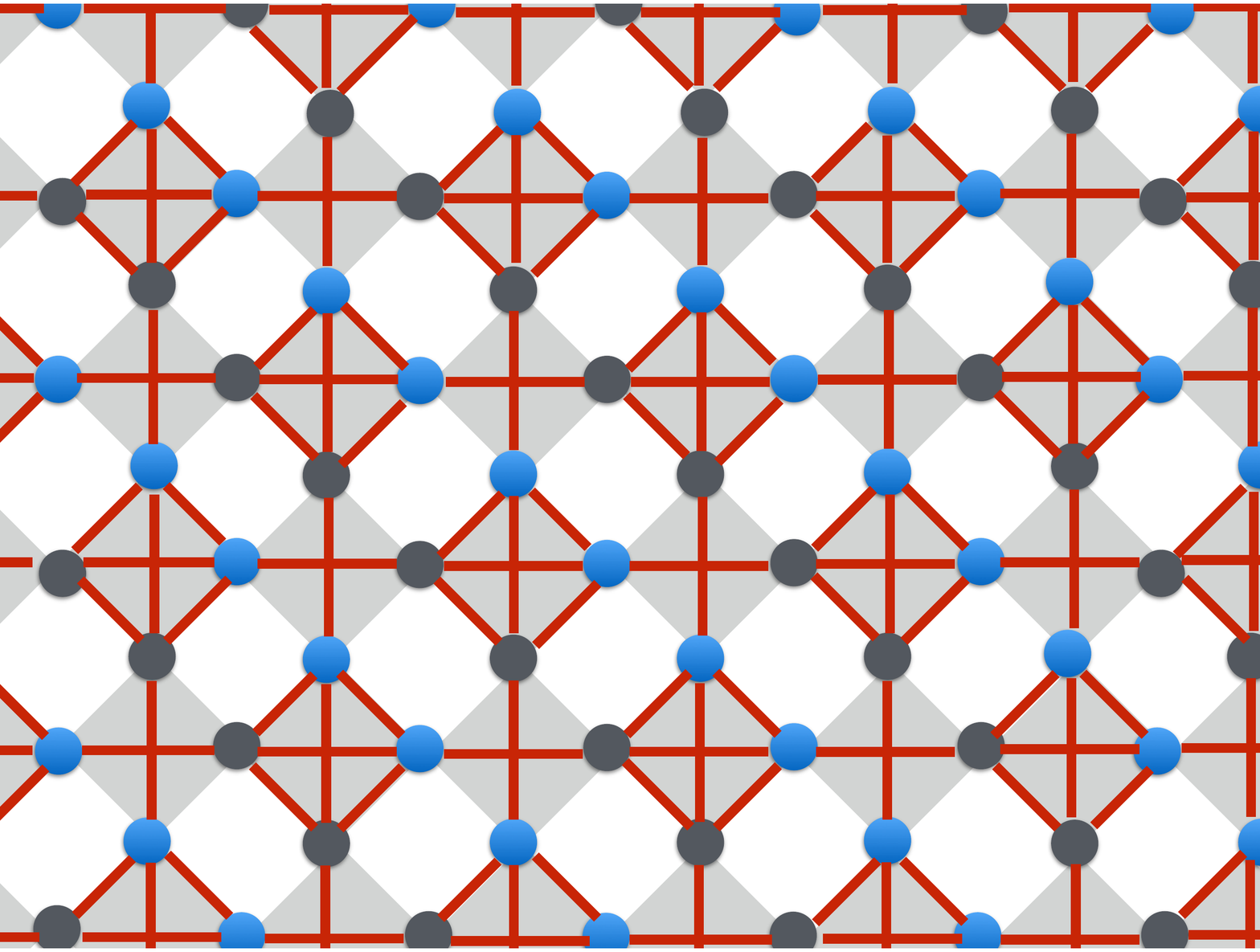}\label{FigGrid_HA1}}
\hfill
\subfigure[\ ]{\includegraphics[width = 0.24\linewidth]{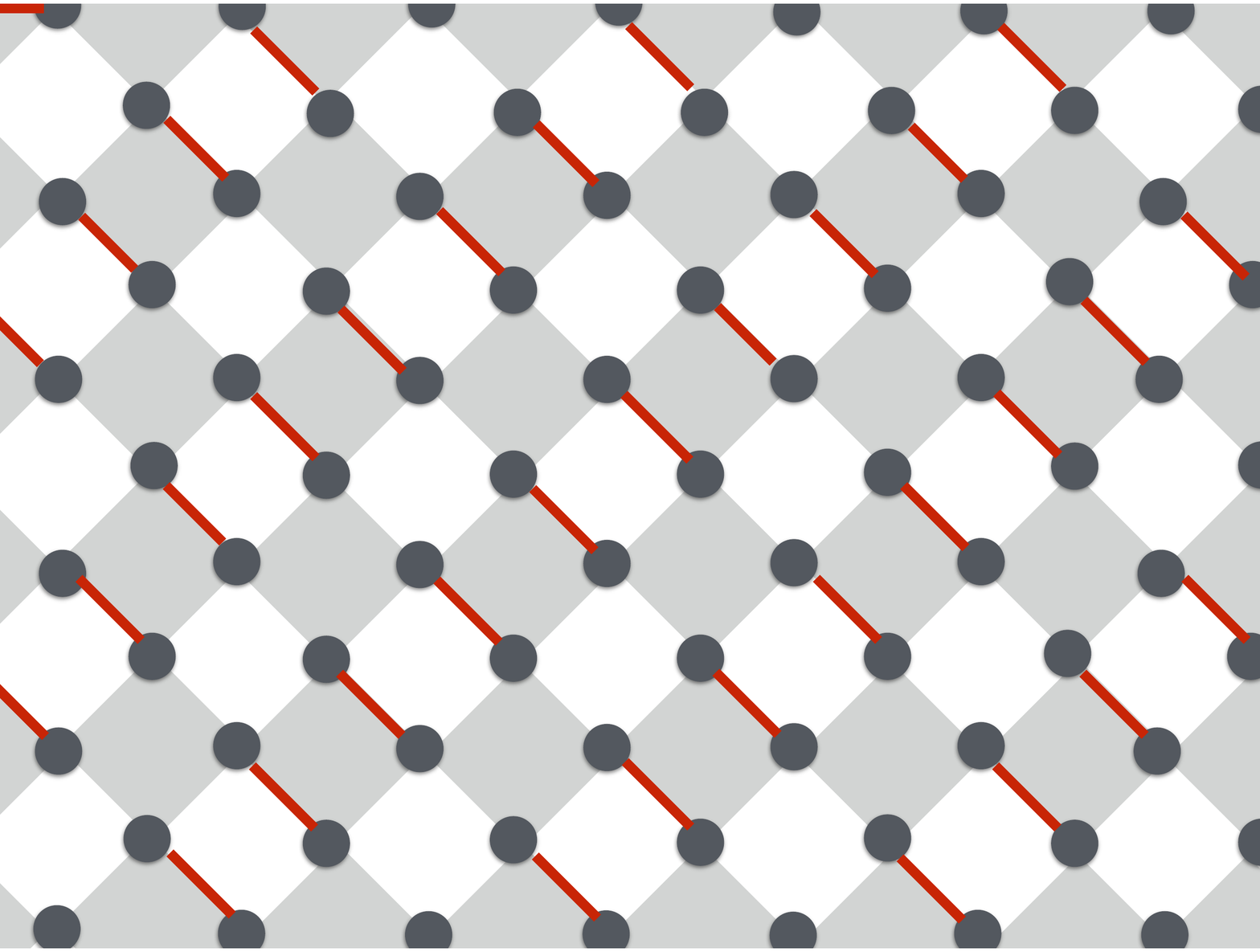}\label{FigGrid_HA2}}
\hfill
\subfigure[\ ]{\includegraphics[width = 0.24\linewidth]{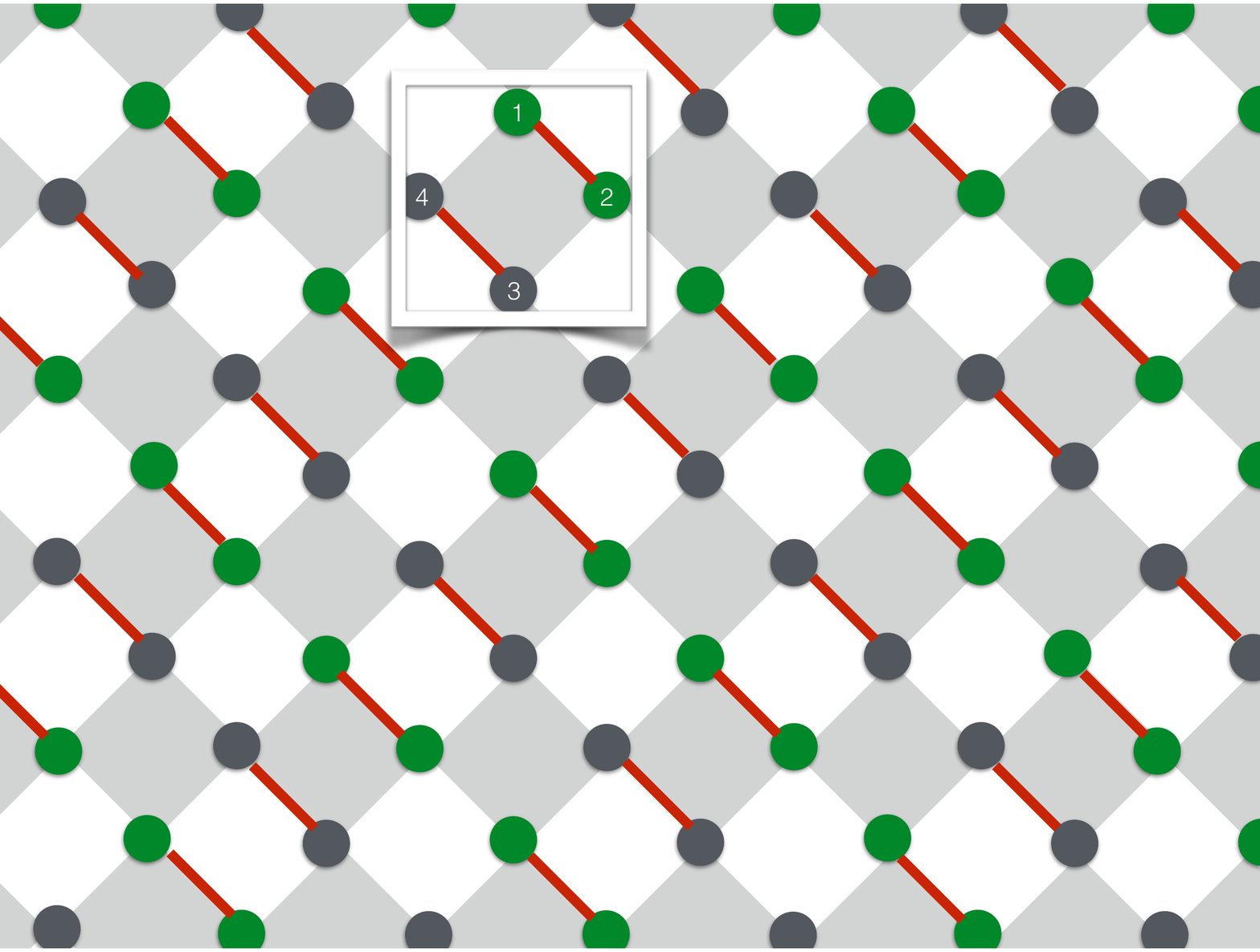}\label{FigGrid_HA3}}
\hspace*{\fill}
\caption{Generating $H_a$ on the grid. (a) Evolving under $H_{\mathrm{temp}}$ [Fig.~\ref{FigGrid_Htemp}], we apply $Z$ pulses on the blue qubits at times $\delta t$ and $2 \delta t$. (b) The effective Hamiltonian at time $2 \delta t$. (c) $H_a$ is generated by conjugating the Hamiltonian in Fig.~\ref{FigGrid_HA2} with Hadamard pulses. The effective Hamiltonian becomes patterns of $Z_1 Z_2+X_3 X_4$, depicted as green coupled pairs ($ZZ$) and black coupled pairs ($XX$).}
\label{fig:HA}
\end{figure*}

\begin{figure*}[t]
\subfigure[\ ]{\includegraphics[width = 0.24\linewidth]{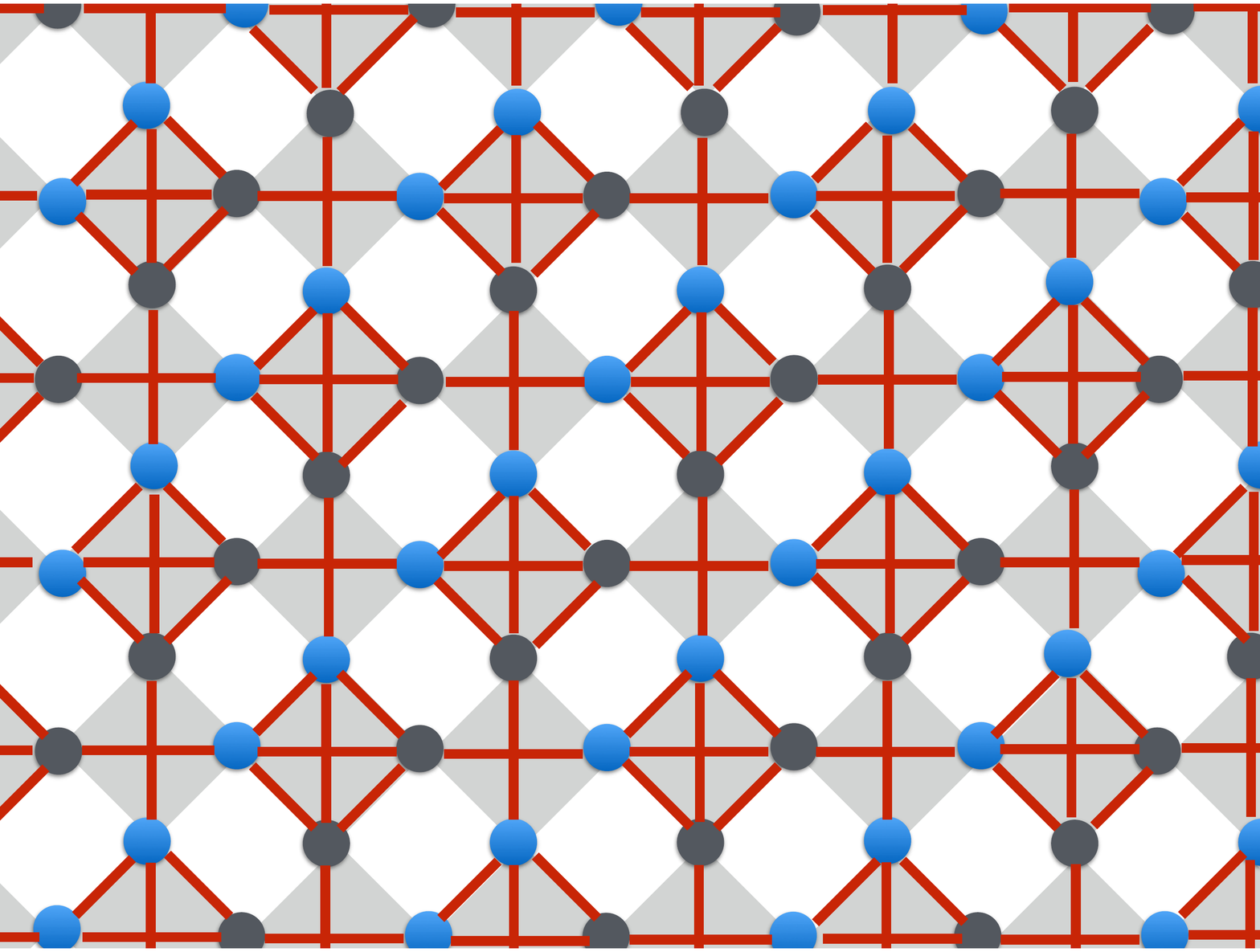}\label{FigGrid_HB1}}
\subfigure[\ ]{\includegraphics[width = 0.24\linewidth]{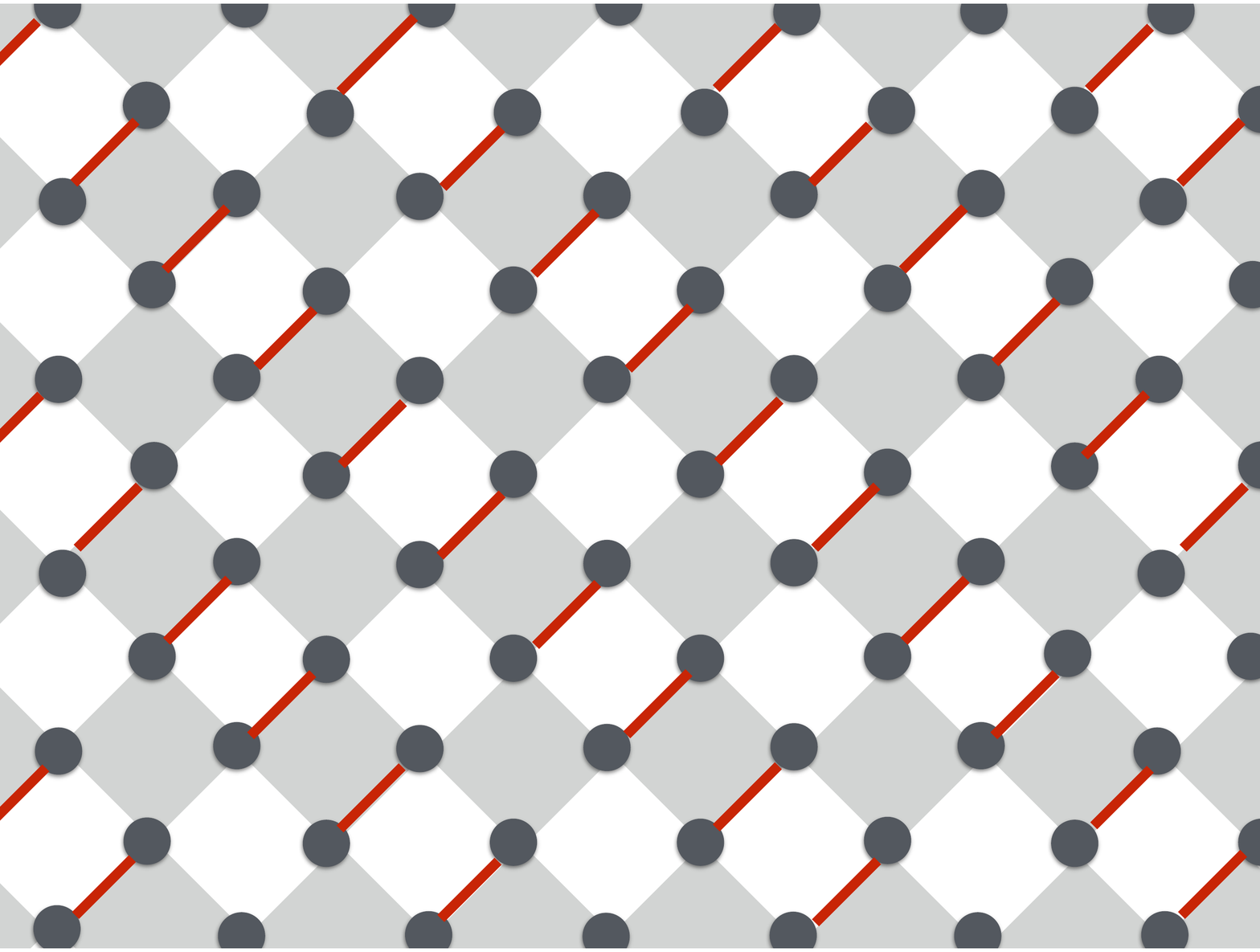}\label{FigGrid_HB2}}
\subfigure[\ ]{\includegraphics[width = 0.24\linewidth]{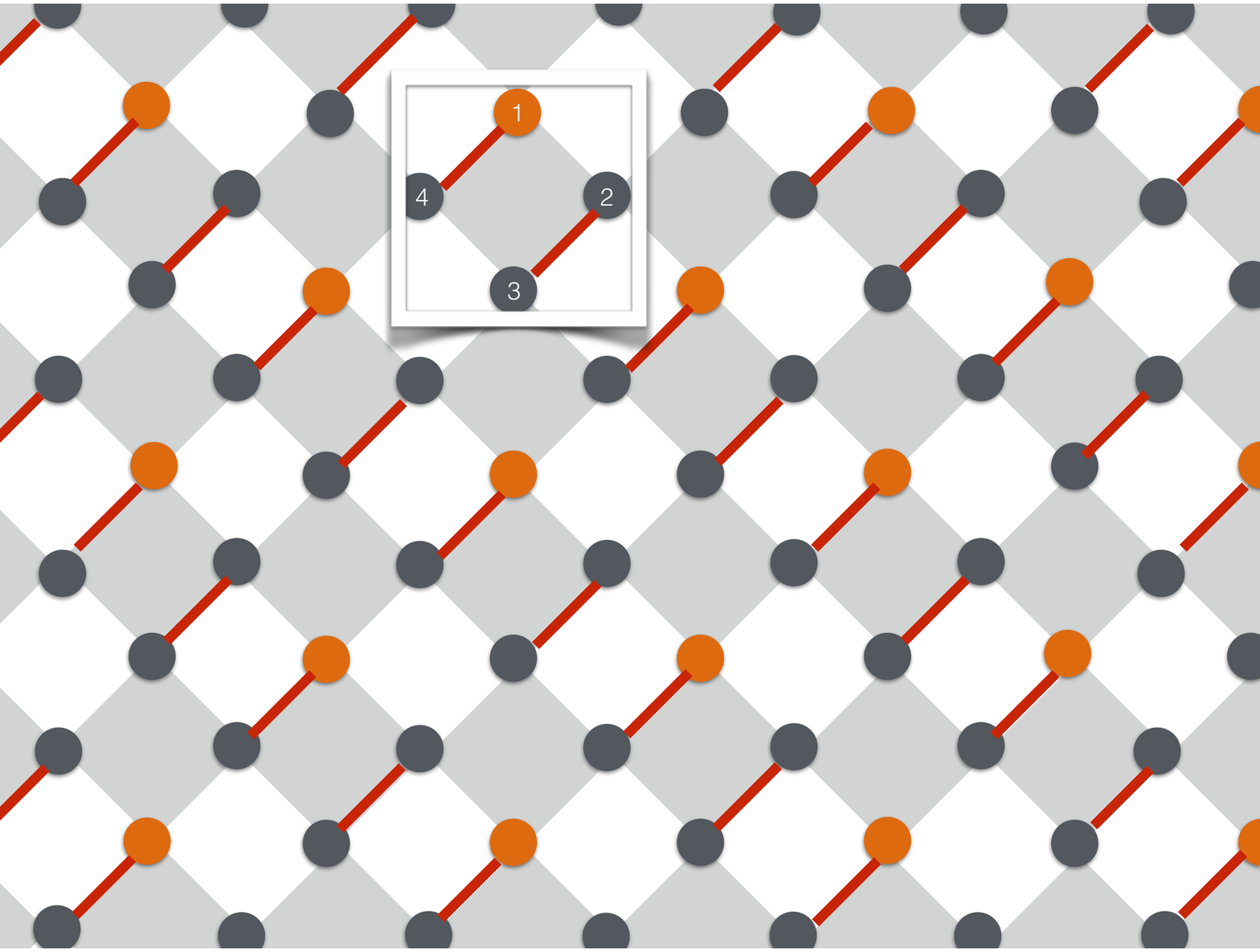}\label{FigGrid_HB3}}
\subfigure[\ ]{\includegraphics[width = 0.24\linewidth]{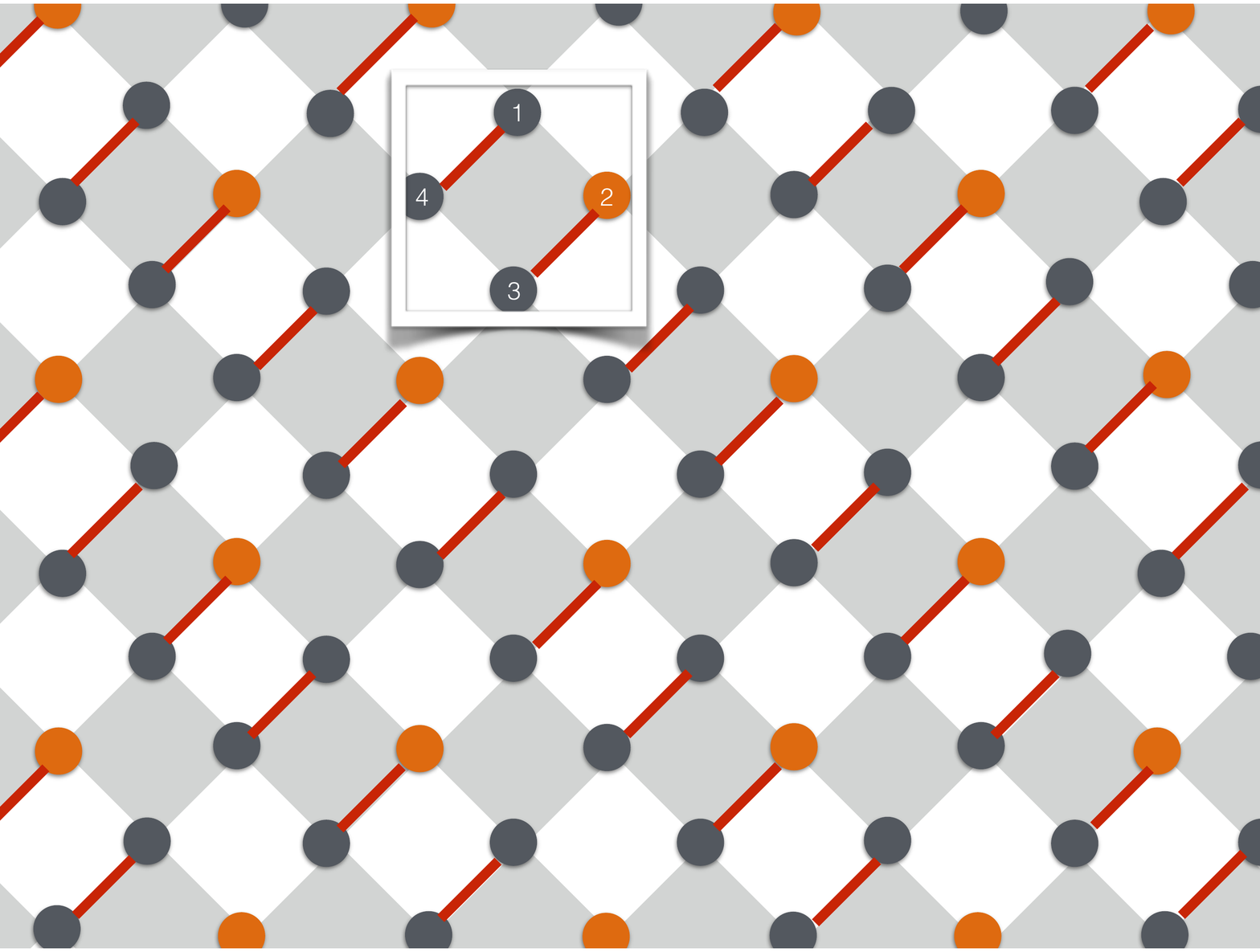}\label{FigGrid_HC}}
\caption{Generating $H_b$ and $H_c$ on the grid. (a) Evolving under $H_{\mathrm{temp}}$ [Fig.~\ref{FigGrid_Htemp}], we apply $Z$ pulses on the blue qubits at times $\delta t$ and $2 \delta t$. (b) The effective Hamiltonian at time $2 \delta t$. (c) $H_b$ is generated by conjugating the Hamiltonian in Fig.~\ref{FigGrid_HB2} with $S$ pulses on the orange qubits. The effective Hamiltonian becomes patterns of $Y_1 X_4+X_2 X_3$, depicted as orange-black coupled pairs ($YX$) and black coupled pairs ($XX$). (d) $H_c$ is generated by conjugating the Hamiltonian in Fig.~\ref{FigGrid_HB2} with $S$ pulses on the orange qubits. Now the effective Hamiltonian becomes patterns of $X_1 X_4+Y_2 X_3$, depicted as black coupled pairs ($XX$) and orange-black coupled pairs ($YX$).}
\label{fig:HBHC}
\end{figure*}

We start by creating patterns similar to $H_x$ by letting $H_X$ evolve and apply $Z$ pulses on the qubits that are colored in blue in Fig.~\ref{FigGrid_Htemp_Zpulses} at times $\delta t$ and $2 \delta t$, but leave alone the qubits that are colored in black. (More precisely, we conjugate by $Z$ pulses all the qubits of a plaquette if the plaquette is in both an odd row and an odd column of the grid, counting from the top left corner.) Every pair of neighboring blue and black qubits is thus decoupled, since $Z$ anticommutes with $XX$, while every pair of neighboring blue qubits is unaffected, since $ZZ$ commutes with $XX$. This results in an effective Hamiltonian, called $H_{\mathrm{temp}}$, shown in Fig.~\ref{FigGrid_Htemp} by red lines as the surviving $XX$ interactions (the plaquettes in an odd row and an odd column, or plaquettes in an even row and an even column of the grid.)
Now we use $H_{\mathrm{temp}}$ and apply pulses similar to Sec.~\ref{Sec:SimOneOperator} on the labelled qubits in Fig.~\ref{FigGrid_Htemp_Labeld} in parallel to generate plaquette operators in half of the grid, as shown in Fig.~\ref{FigGrid_4Body}. The steps of this procedure, and the corresponding components $H_{\{a,b,c\}}$ are presented in Figs.~\ref{fig:HA} and \ref{fig:HBHC}.

As discussed in Sec.~\ref{Sec:SimOneOperator}, by applying $Z_1 Y_2$ (but now in parallel on the whole grid) one can remove all the extra unwanted terms, and all that remains are the plaquette operators shown in Fig.~\ref{FigGrid_4Body}.

Up to now, the effective Hamiltonian only contains half of the plaquette operators needed. The other half can be generated by repeating the same procedure, but shifting the locations of the pulses by one qubit. In this case plaquette operators in the shaded area of Fig.~\ref{FigGrid_4Body} are generated. The combined process gives all the plaquette operators needed:
\begin{eqnarray}
e^{-ih \delta t^3 \sum_{v}{A_v}}+O(\delta t^4)\ ,
\end{eqnarray}
where $h\delta t^2$ is strength of the effective Hamiltonian, and $h$ accounts for various numerical factors (such as $64/N_X$ in Sec.~\ref{Sec:SimOneOperator}).

\subsection{Simulating all the vertex operators on the grid} 
Noticing, using Eq.~\eqref{eq:conjugate-by-u}, that
\begin{eqnarray}
W_1 W_2 W_3 W_4 e^{-i X_1 X_2 X_3 X_4 \delta t} W_1 W_2 W_3 W_4 \nonumber \\
=e^{-i Z_1 Z_2 Z_3 Z_4 \delta t}\ ,	
\end{eqnarray}
the process to create the vertex operators on the grid is similar to the process for plaquette operators. One can just shift the location of pulses and then conjugate the whole effective Hamiltonian with Hadamard pulses on all the qubits of the grid. 

\subsection{Strength of the effective Hamiltonian} \label{Sec:effectiveHamiltonain}

By simulating the plaquette operators and the vertex operators consecutively, we can effectively simulate the toric code Hamiltonian in $N_{\rm sim}=320$ steps,\footnote{$10*2$ comes from the commutator method, a factor of $4$ comes from generating each of $H_{\{a,b,c\}}$ from the original Hamiltonian $H_X$. A factor of $2$ is needed to generate the other half of the plaquettes, and another factor of $2$ is needed to generate the vertex operators. Thus $N_{\rm sim}=10*2*4*2*2=320$.}
and so in time $N_{\rm sim} \delta t$, independent of size of the grid. 

The generated $4$-body operator on the grid is $2^9 \delta t^3 H_p$. Taking the strength of $H_X$ as $1$, the strength of each $H_{\{a,b,c\}}$ is $4$. Inserting these into the commutators, $[[H_a,H_b],H_c]]$ has strength $4^4$. Another factor $2$ is added when we remove the error term. Thus $4^2*2 \delta t^3=2^9 \delta t^3$. That is, we use $H_X$ to generate the toric code Hamiltonian while making it $2^9 \delta t^3/ 320 \delta t=1.6 \delta t^2$ times weaker. As expected, decreasing  $\delta t$ makes the ratio smaller.

\subsection{Boundaries and the surface code} 
\label{Sec:boundries}

The toric code is defined on a torus, without boundaries. We are also interested in the surface code, which has boundaries~\cite{Bravyi:1998xy}. To simulate the boundaries for the surface code Hamiltonian, whether outer boundaries or the inner cuts representing qubits, two other procedures are also needed. The first is generating holes (either $Z$-cut or $X$-cut), which is straightforward. For example, wherever it is necessary to create a hole one can choose $H_c'$ such that it commutes with $[H_a,H_b]$. To accomplish this, rather than applying $S_2$ in Eq.~\eqref{eq:generateHc}, if we apply the Hadamard pulse $W_2$, then $H_c'=Z_2 X_3+X_1 X_4$, which commutes with $[ H_a,H_b]$, and so the effective Hamiltonian is zero to order $\delta t^3$. Thus, simply by using a Hadamard pulse rather than phase gate on a specific qubit we can generate a hole.

The second procedure needed is the simulation of $3$-body interactions. One approach is to simulate them using Eq.~\eqref{eq:2commutatormethod} at order $\delta t^2$, while eliminating errors of order $\delta t^3$. Using the same pattern as Sec. \ref{sec:SimulatingAllPlaquettes}, starting from $H_x$ in Fig.~\ref{fig:OnePlaquette} we can choose $H_a=Z_1 X_2+ X_3 X_4$ and $H_b=Y_1 X_4 +X_2 X_3$. This results in an effective Hamiltonian $X_1 X_2 X_4 \delta t^2 + (Y_1 X_4+Z_1 X_2) \delta t^3 +O(\delta t^4)$. By decoupling using one extra pulse $X_1$, we can remove the terms proportional to $\delta t^3$. Also, if in Fig.~\ref{fig:OnePlaquette} the third qubit and the terms it involves are missing (these are edges of the Hamiltonian, corresponding to outer boundaries), then again we can generate $H_a=Z_1 X_2$ and $H_b=Y_1 X_4$, and then apply $X_1$ with the same final result.
 
To balance the different strength of the generated terms (boundaries are proportional to $\delta t^2$ while the rest of the construction is proportional to $\delta t^3$), the proportion of time slices implemented from each of these terms (in Trotterization of the evolutions) can be chosen to be $\delta t$. Another approach is to implement these boundaries together with the other terms, but with flipping the signs of some of the Hamiltonians so that all but a fraction $\delta t$ of the 3-body terms cancel out. (In our example, we could use $Z_1$ pulses: $Z_1 e^{-i X_1 X_2 X_4 \delta t^2} Z_1= e^{i X_1 X_2 X_4 \delta t^2}$.)

An alternative approach to avoid the complicated scheduling described above is to apply pulses such as $e^{-i Y \delta t}$ to implement single-body Hamiltonians. For example, starting from $H_x$ in Fig.~\ref{fig:OnePlaquette} we can repeat the procedure in  Sec. \ref{Sec:SimOneOperator} to generate $H_a=Z_1 X_2+ X_3 X_4$ and $H_b=Y_1 X_4 +X_2 X_3$, but now generate $H_c= X_1 X_4+ X_3 X_4 +X_1 X_3+ Y_2$. If so, the effective Hamiltonian to order $\delta t^3$ (after applying $Z_1 Y_2$ pulses to remove the extra terms, as mentioned in  Sec. \ref{Sec:SimOneOperator} ) would become $X_1 X_2 X_4$ as desired. To generate such an $H_c$ we can follow this procedure:
\begin{eqnarray}
Z_2 e^{-i H_x \delta t} Z_2 &&e^{-i H_x \delta t} e^{-i 2 Y_2 \delta t}\nonumber \\
&&=e^{-i 2 (X_1 X_4+ X_3 X_4 +X_1 X_3+ Y_2) \delta t}.
\end{eqnarray}

\subsection{$\pi/4$-Conjugation Method}

An alternative method for simulating high-weight Hamiltonians is based on the {$\pi/4$-conjugation} identity (see, e.g., \cite{lidar2001decoherence}). This identity for Pauli operators $A,B$ with $\{A,B\}=0$ is
\begin{eqnarray}
e^{-i\pi/4 A}e^{i \theta B}e^{i\pi/4 A}=e^{i\theta (iAB)},
\end{eqnarray}
which is exact. 
Obviously if $[A,B]=0$ then we have
\begin{eqnarray}
e^{-i\pi/4 A}e^{i \theta B}e^{i\pi/4 A}=e^{i\theta B}.
\end{eqnarray}

Assume the entangling Hamiltonian is $c H_x=c \sum_{i<j}{X_iX_j}$ with $c$ being a constant representing the strength of the entangling Hamiltonian:
\begin{eqnarray}
e^{-i \frac{\pi}{4c} (cH_x)} e^{-i \theta Y_1} e^{i \frac{\pi}{4c} (cH_x)}= e^{i \theta Z_1 X_2 X_3 X_4}.
\end{eqnarray}
Using this we can show that:
\begin{eqnarray}
&&W_1 [ e^{-i \frac{\pi}{4c} (cH_e)} e^{i \Delta t Y_1}
(Y_1  e^{-i \frac{\pi}{4c} (cH_e)} Y_1)]  W_1 \nonumber \\
&&=i e^{-i \Delta t X_1 X_2 X_3 X_4}.
\end{eqnarray}
So we apply three fast pulses while having $H_x$ on for a length of time $2\pi/4c$. It is important to notice that in this method the simulation time, here $2 \pi/4c$, is independent of $\Delta t$, the time over which we wish to simulate the evolution.

To simulate plaquette operators we again can conjugate this with Hadamard pulses:
\begin{eqnarray} 
W_1 W_2 W_3 W_4  e^{-i \Delta t X_1 X_2 X_3 X_4} W_1 W_2 W_3 W_4 \nonumber \\
=e^{-i \Delta t Z_1 Z_2 Z_3 Z_4}.
\end{eqnarray}

Implementation of the Hamiltonian on the grid can be done similarly to Sec.~\ref{sec:SimulatingAllPlaquettes}, with slight modifications.

Using this procedure there is no error in simulation (as the identity is exact)  and it takes a time proportional to $\pi/c$ independent of $\Delta t$.


{In contrast to the commutator method, where using the entangling Hamiltonian for a short time produces a weak effective Hamiltonian, here the entangling Hamiltonian has to evolve for the fixed time  $\pi/4c$ to produce the desired effective Hamiltonian. For an experimental setup that can only generate weak entangling Hamiltonian (small $c$), this method requires a large time which may make the implementation more prone to noise. For experiments with access to strong entangling Hamiltonians compared to the single-body gates and measurements, this method is beneficial as it is exact and the simulation time is independent of the time over which we wish to simulate a Hamiltonian. See \cite{dur_quantum_2008} for a comparison of the $\pi/4$-conjugation method to the commutator method in a Markovian environment.}

\subsection{Applying the scheme to fault-tolerant holonomic quantum computation in surface codes} 
\label{sec:faulttoletant}


Using adiabatic deformation of the surface code Hamiltonian in combination with active error correction it is possible to achieve fault-tolerant universal quantum computation \cite{zheng_fault2015}. In such a scheme, a gapped Hamiltonian protects the ground space from the deleterious effects of thermal noise, and the gates are applied by slowly deforming the Hamiltonian, such that the desired gate is applied to the ground state of the new Hamiltonian (holonomic quantum computation \cite{zanardi1999holonomic}). Explicit constructions to implement all the  gates needed for universal quantum computation were proposed in \cite{zheng_fault2015} for such deformations.

The Hamiltonians needed for these schemes can be made geometrically local \cite{zheng_fault2015}, but in practice it is challenging to implement them as they consist of up to $4$-local interactions.  Using the methods we proposed above, it is straightforward to simulate these Hamiltonians as well. For the purpose of illustration we choose a few examples from \cite{zheng_fault2015} and explain the needed procedure, but the constructions for the remaining needed interactions are similar.

As the first example, we note that the creation of a $\ket{+}$ state for an $X$-cut double qubit (and of a $\ket{0}$ state for a $Z$-cut double qubit) is done by just turning off two stabilizer terms (Sec IV.A in \cite{zheng_fault2015}). As explained in Sec.~\ref{Sec:boundries}, this can be done by changing a pulse such that the $O(\delta t^3)$ effective Hamiltonian vanishes.

We also need the ability to enlarge a hole. The proposed time-dependent Hamiltonian implementing the deformation is of the form:
\begin{equation}
H(t)=-J\left((1-\frac{t}{t_1}) B_2 + \frac{t}{t_1} X_1 + \sum_{p \neq 1,2}{B_p} + \sum_{v}{A_v}\right),	
\end{equation}
for $t \in [0,t_1]$ (see Eq.~(37) in \cite{zheng_fault2015}, with a specific choice of the monotonic function). 

To simulate the time-dependent Hamiltonian, we can use Trotter-Suzuki expansions  (see, e.g., \cite{poulin_quantum_2011}), and simulate each piece by applying the appropriate pulses. To do so, we define $\Delta t=\frac{t_1}{N_{tr}}$ and approximate the evolution generated by the time-dependent Hamiltonian with  ordered evolutions consisting of several time independent Hamiltonians:
\begin{eqnarray}
\prod_{m=0}^{N_{tr}} e^{-i ((1-\frac{m}{N_{tr}})  B_2 + \sum_{p \neq 1,2}{B_p}) J \Delta t} \nonumber \\
 e^{-i \frac{m}{N_{tr}} X_1 J \Delta t} e^{-i J \sum_{v}{A_v} \Delta t} 
\end{eqnarray}

To simulate these terms, one can generate the weakest 4-body interaction in this expansion using the building blocks we constructed  in Sec.~\ref{sec:SimulatingAllPlaquettes}, setting $\frac{J \Delta t}{N_{tr}}=h \delta t^3$:
\begin{eqnarray}
\prod_{m=0}^{N_{tr}} (e^{-i B_2 h \delta t^3})^{N_{tr}-m} (e^{-i \sum_{p \neq 1,2}{ B_p h \delta t^3}  })^ {N_{tr}} \quad \nonumber \\
(e^{-i X_1 h \delta t^3})^m (e^{-i h \sum_{v}{A_v} \delta t^3})^{N_{tr}}. 
\end{eqnarray}
Each 4-body term in this expansion can be simulated as described in Sec.~\ref{sec:SimulatingAllPlaquettes}. We notice that $B_2$ can be constructed in parallel with the rest of the plaquette operators. Also, $e^{-i X_1 h \delta t^3}$ is just a single qubit rotation applied using a pulse.

The error in the Trotterization procedure scales as $\Delta t^2$, and is independent of the system size (the only non-commuting term is $X_1$; see \cite{poulin_quantum_2011}).

The same procedure works for the other  Hamiltonians needed, such as another Hamiltonian needed to enlarge a hole (Eq.~(39) in \cite{zheng_fault2015}), or the Hamiltonian needed for moving logical qubits (Eq.~(47) in \cite{zheng_fault2015}).

The
$\bar{Z}$ and $\bar{X}$ logical operators of the surface code are just strings of $Z$ and $X$ operators that connect appropriate boundaries (or holes). Thus any product thereof can be formed by applying parallel $Z$ and $X$ pulses (in practice one may avoid applying them and instead keep track of updates in software).

\section{Effect of noise} \label{sec:noise}
So far we have considered only the noise-free case. In this section we study the effect of noise on the simulation, and propose methods to suppress the effective noise of this process. 

Again we use the entangling Hamiltonian $H_X$ [Eq.~\eqref{eq:H_X}], but now we assume the presence of a general non-Markovian bath with Hamiltonian $H_B$, interacting with the system via the Hamiltonian $H_{SB}$. For now we assume that this interaction Hamiltonian is $1$-local:
\begin{eqnarray}
H_B&=& I \otimes B_0, \\
H_{SB}&=&\sum_{i}\sum_{\alpha \in\{x,y,z\}}{\sigma_i^\alpha \otimes B_i^{\alpha}}\ ,
\end{eqnarray}
where $i$ labels the the system qubit (when it is clear from the context, we will combine this index and the Pauli operator index into a single index: $H_{SB}=\sum_{\beta}{\sigma_\beta \otimes B_{\beta}}$.) The total Hamiltonian is:
\begin{eqnarray} \label{eq:totalHamiltonian}
H=H_X+H_B + \lambda H_{SB}\ ,
\end{eqnarray}
with the dimensionless parameter $\lambda$ being the strength of the system-bath coupling. 

As the effective $4$-body interactions constructed in previous sections appear in order $\delta t^3$, our strategy is to choose a series of DD pulses that commute with the Hamiltonian $H_X$ but suppress the errors. This use of DD pulses that commute with the Hamiltonian is an example of the general idea of using the stabilizer generators of a quantum error correction (or detection) code to perform a decouple-while-compute operation, which is possible since the Hamiltonian is a sum of the logical elements of the same code \cite{Paz-Silva:2013tt,KhodjastehLidar:08,West:10}. We will show how to protect the Hamiltonian $H_X$ against the noise at least up to order $\delta t^3$, so that the simulated Hamiltonian is not overwhelmed by the noise. To simulate the desired interactions, we combine the pulses designed for simulation with these DD pulses.

Note that DD schemes are effective when the timescale of the bath is long compared to the timescales of the pulses, and in fact DD can be shown to fail in the Markovian limit \cite{DDMarkovian}. Therefore the method proposed here is only effective for non-Markovian environments. 

We begin by first finding pulses that can suppress the errors in a general non-Markovian environment, and then discuss how further assumptions on the locality of the bath and the system-bath interaction allow for more efficient schemes. 

To analyze the effects of DD pulses, and the corresponding effective Hamiltonian, it is handy to use the Magnus expansion (see Appendix \ref{app:Magnus}.) Here we choose the simplest DD schemes to illustrate the main ideas, but there is much room to use more sophisticated DD schemes (see, e.g., Ref.~\cite{lidar2013quantum} for a review).

\subsection{Protecting against noise at first and second order in $\delta t$}
To remove the first order noise terms, we conjugate the Hamiltonian $H_X$ with pulses $I^{\otimes N}, X^{\otimes N},Y^{\otimes N},Z^{\otimes N}$ in $4$ steps. These pulses all commute with the Hamiltonian but remove the noise from the effective Hamiltonian to first order, via symmetrization \cite{zanardi_symmetrizing_1999,viola_universal_1999}. 

Then, using the fact that all the even-order terms of the Magnus expansion vanish for time-symmetric Hamiltonians \cite{ng_combining_2011}, we simply append another $4$ steps, conjugating with the same $4$ pulses but in reverse order. This removes the first and second order terms of the noise from the effective Hamiltonian:
\begin{eqnarray} 
U_{sec}&=&(I^{\otimes N} e^{-i H \delta t} I^{\otimes N})(X^{\otimes N} e^{-i H \delta t} X^{\otimes N}) \nonumber \\
&&(Y^{\otimes N} e^{-i H \delta t} Y^{\otimes N})(Z^{\otimes N} e^{-i H \delta t} Z^{\otimes N})\nonumber \\
&&(Z^{\otimes N} e^{-i H \delta t} Z^{\otimes N}) (Y^{\otimes N} e^{-i H \delta t} Y^{\otimes N}) \nonumber\\
&&(X^{\otimes N} e^{-i H \delta t} X^{\otimes N})(I^{\otimes N} e^{-i H \delta t} I^{\otimes N}) \nonumber \\
&=&e^{-i (H_X+H_B) 8 \delta t +O(\delta t^3)}\ .
\end{eqnarray}
Note that the middle $Z^{\otimes N}$ terms cancel, and that we can simplify this sequence to:
\begin{eqnarray} 
U_{sec}&=& e^{-i H \delta t} X^{\otimes N} e^{-i H \delta t} Z^{\otimes N}  e^{-i H \delta t} X^{\otimes N} e^{-2i H \delta t} \nonumber \\
&& X^{\otimes N} e^{-i H \delta t} X^{\otimes N} e^{-i H \delta t} X^{\otimes N} e^{-i H \delta t}  \nonumber \\
&=&e^{-i (H_X+H_B) 8 \delta t +O(\delta t^3)}\ .
\label{eq:Usec}
\end{eqnarray}

At this point, having evolved for $8\delta t$ (independent of the grid size), we have protected the $H_X$ Hamiltonian from the effect of noise to first and second order in $\delta t$.

\subsection{Third order error terms} \label{app:TwirlingProt}

First using a lemma we describe a general approach to protect a Hamiltonian while averaging out all other errors at a fixed order of time. It can be understood as an application of the symmetrization schemes proposed in \cite{viola_dynamical_2000,zanardi_symmetrizing_1999}.

\begin{mylemma} \emph{(Symmetrization lemma: Protecting interaction)}
Let $P_1,\ldots,P_n$ be commuting Pauli operators, and let $\mathcal{P}$ be the group generated by these Pauli operators. Denote the set of all Pauli operators that commute with all the $P_i$s by $\mathcal{N}(\mathcal{P})$ {(the normalizer of $\mathcal{P}$)}. Then for any Pauli operator $s$
\begin{eqnarray}
&&\forall s \notin \mathcal{P:} \sum_{g\in \mathcal{\mathcal{N}(\mathcal{P})}}{g s g^{\dagger}}=0,\\
&&\forall s \in \mathcal{P:} \sum_{g\in \mathcal{\mathcal{N}(\mathcal{P})}}{g s g^{\dagger}}=s |\mathcal{\mathcal{N}(\mathcal{P})}|.
\end{eqnarray} 
\end{mylemma}

As a special case, if $\mathcal{P}$  only includes the identity operator we recover the usual Pauli twirling lemma (see, e.g.,  \cite{wilde2013quantum}.)
\begin{proof}
See Appendix \ref{app:Twirling}.
\end{proof}

Using this lemma for the particular plaquette operator discussed in section \ref{Sec:SimOneOperator}, choosing $P_1=X_1X_2$,$P_2=X_2 X_3,P_3=X_3X_4$ we need  $4^4 / 2^3 = 32$ pulses to cancel out all the third-order errors. (Again, by symmetrization and doubling the number of pulses we get can get rid of fourth-order error as well.)

But to protect the whole grid, without additional assumptions about the locality of the bath part of the interaction Hamiltonians (the $\{B_\alpha\}$ operators), the number of pulses needed to cancel the terms at third order may grow with the grid size. The reason is that at higher orders, multiqubit error terms begin to appear without a locality assumption.  To average out these terms by DD methods, the number of pulses must grow with the number of qubits (see Appendix \ref{app:DDMultiqubitGrows}).

Later we will consider the case where the terms $\{B_\alpha\}$ are geometrically local, but here we show that even without this assumption it is possible to protect the Hamiltonian against the dominant noise terms appearing at third order in $\delta t$ using a few pulses. To see this, we note that at order $\delta t^3$, the terms that are first order in the coupling strength $\lambda$ have a specific form: they are nested commutators of $\sigma_i^\alpha \otimes B_i^{\alpha}$ with two $H_X$ terms:
\begin{eqnarray}
&&[[H_X,\sigma_i^\alpha \otimes B_i^{\alpha}],H_X] \nonumber \\
&&=[[X_i \otimes \sum_{{\langle\langle i,i'\rangle\rangle}} X_{i'}, \sigma_i^\alpha \otimes B_i^{\alpha}],H_X] \nonumber \\
&&=[ [X_i,\sigma_i^\alpha] \otimes \sum_{{\langle\langle i,i'\rangle\rangle}} X_{i'} \otimes B_i^{\alpha}, H_X] \nonumber \\
&&=[[X_i,\sigma_i^\alpha],X_i] \otimes (\sum_{{\langle\langle i,i'\rangle\rangle}} X_{i'})^2 \otimes B_i^{\alpha}.
\end{eqnarray}




These terms are either $Y_{i}, Z_{i}, Y_{i} X_{i'} X_{i''}$ or $Z_{i} X_{i'} X_{i''}$, and are thus decoupled away by conjugating the evolution in Eq.~\eqref{eq:Usec} with $I^{\otimes N}$ and $X^{\otimes N}$ pulses:
\begin{eqnarray}
(I^{\otimes N}U_{sec}&&I^{\otimes N})(X^{\otimes N}U_{sec}X^{\otimes N}) \nonumber\\
&&=e^{-i (H_X+H_B) 16 \delta t +O(\lambda^2 \delta t^3)+ O(\delta t^4)} 
\end{eqnarray}
In fact, these two pulses not only remove the $\lambda^1$  terms at order $\delta t^3$, but at any order of $\delta t$. 
This is expected, as the first-order terms in $\lambda$ are the result of the system Hamiltonian spreading the noise across more qubits. As the system Hamiltonian terms are all local and commuting, the noise to first order in $\lambda$ can only have the form of either $Y_{i}X \dots X$ or $Z_{i} X \dots X $, and so conjugating with the proposed pulses can remove them.

Assuming that the system is weakly coupled to the environment (small $\lambda$), the dominant terms in the expansions are the lower powers of $\lambda$, so using just two additional pulses can cancel out all errors to first order in $\lambda$.

To sum up, we obtain this DD protection for $H_X$ on the grid by evolving for time $16\delta t$, without any assumptions about geometric locality of the interaction Hamiltonian. With the noise removed by this DD procedure to order $\delta t^3$, we can use our earlier constructions to produce the effective Hamiltonian without its being overwhelmed by noise.

\subsection{Strength of the simulated Hamiltonian vs. effective noise}
The effective noise process on the system depends on the particular schemes used for simulation and for DD.  
The ratio between this effective noise strength and the strength of the simulated Hamiltonian is important. If this ratio is small enough, the energy gap protection of the simulated Hamiltonian can be effective against the effective noise. Here we give estimates of the strength of each of these terms, but we also expect that for specific systems with more knowledge about the form of the noise, better bounds can be achieved.

\begin{figure}
\centering
\includegraphics[width=1\columnwidth]{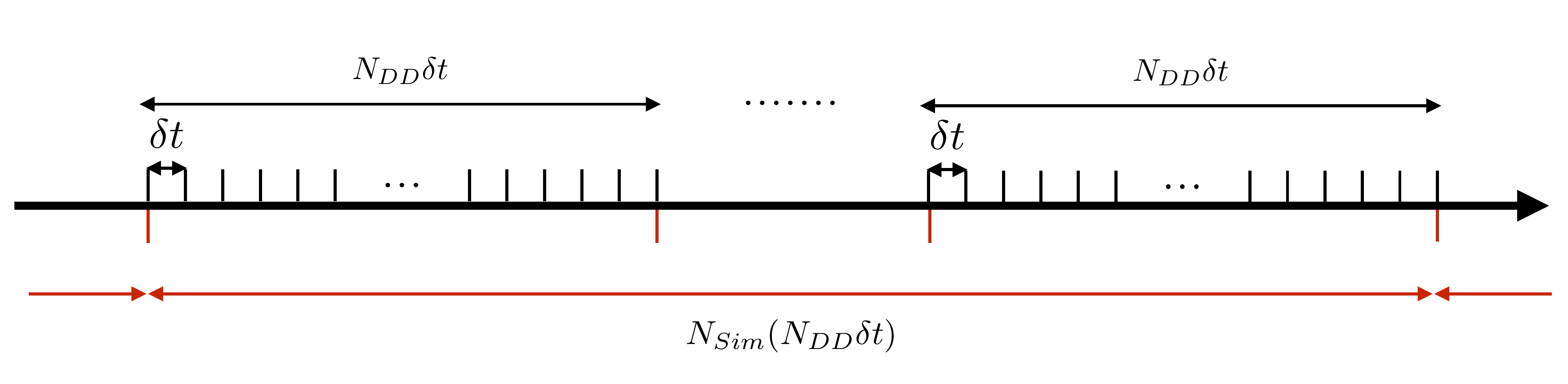}
\caption{{Timing of the pulses: combining a DD scheme with $N_{\rm DD}$ pulses and a simulation with $N_{\rm sim}$ steps.}}
\label{FigPulses}
\end{figure}

We assume $\delta t$ is the time interval between pulses. This is most likely determined by the physical limitations of the experiment. As described earlier, we choose $N_{\rm DD} N_{\rm sim}$ pulses to simulate a building block of the desired Hamiltonian at order $\delta t^3$, with the effective noise at second order in $\lambda$ and third order in $\delta t$ (Fig.~\ref{FigPulses}).

We first find a bound on the effective noise strength in each DD interval, and then add the effect of simulation on it. Following \cite{ng_combining_2011}, we define the effective noise strength as:
\begin{eqnarray}
\eta=\norm{U(N_{\rm DD} \delta t)- U_{\rm ideal}(\delta t)},
\end{eqnarray} 
with $U$ being the unitary generated by $H_X+H_B+H_{SB}$ with the DD pulses applied in between. The operator 
\beq
U_{\rm ideal}(t)\equiv e^{-i (H_B+H_X) N_{\rm DD}  t}
\eeq 
is the ideal case, with $H_{SB}$ absent. We denote the cumulative  unitary generated by all pulses at each time (not including the Hamiltonian terms) by $U_{\rm pulse}(t)$. We assume that the pulses all commute with the system Hamiltonian $H_X$ 
and also assume that the product of all pulses in one cycle gives identity: $U_{\rm pulse}(N_{\rm DD} \delta t)=I$. Moving to the interaction picture defined by the pulses we have:
\begin{eqnarray}
\eta=\norm{\tilde{U}(N_{\rm DD} \delta t) - U_{\rm ideal}(\delta t)}\ ,
\end{eqnarray}
with 
\begin{equation}
\tilde{U}(N_{\rm DD} \delta t)=\Texp\left[-i \int_0^{N_{\rm DD} \delta t}{H_X+H_B+\tilde{H}_{SB}(t) dt}\right]\ ,
\end{equation}
and 
 \begin{eqnarray}
 \tilde{H}_{SB}(t)=U_{\rm pulse}^{\dagger}(t) H_{SB} U_{\rm pulse}(t).
 \end{eqnarray}
The Magnus expansion can be used to approximate $\tilde{U}(N_{\rm DD} \delta t)$ with an effective Hamiltonian $H_{\rm eff}$ up to arbitrary order. Denoting $\tilde{U}(N_{\rm DD} \delta t)=\exp(-iH_{\rm eff} N_{\rm DD} \delta t)$, the Magnus expansion gives each term in $H_{\rm eff}$ (see Appendix~\ref{app:Magnus}):
\begin{eqnarray}
H_{\textrm{eff}}= \sum_{k=0}{H_{\textrm{eff}}^{(k)} } \ .
\end{eqnarray}

Using this we have an upper bound on $\eta$ as 
\begin{eqnarray}
\eta \leq N_{\rm DD} \delta t \norm{\sum_{k=0}{H_{\textrm{eff}}^{(k)}-(H_B+H_X)}}\ .
\end{eqnarray}
Assuming that the pulses achieve complete decoupling for first and second order, we have $H_{\textrm{eff}}^{(0)}=H_B+H_X$ and $H_{\textrm{eff}}^{(1)}=0$. So we have
\begin{eqnarray} \label{Eq:UpperDD}
\eta \leq N_{\rm DD} \delta t \left(\norm{H_{\textrm{eff}}^{(2)}} + \norm{\sum_{k\geq 3}{H_{\textrm{eff}}^{(k)}}}\right)
\end{eqnarray}

Now we can use the bounds on Magnus terms derived in \cite{ng_combining_2011}, but a slight modification is needed. First is the addition of the $H_X$ Hamiltonian, and the other is taking into account the effect of the two extra pulses we used to remove the first order terms in $\lambda$.

For the first term in Eq.~\eqref{Eq:UpperDD}, adding the terms corresponding to $\lambda^2,\lambda^3$ (adding Eqs.~(112) and (113) in \cite{ng_combining_2011}: changing $\beta$ to $\norm{H_B+H_X}$ and replacing $\epsilon$ with $\norm{H_B+H_X+\lambda H_{SB}}$), we have:
\begin{eqnarray}
 &&\norm{H_{\textrm{eff}}^{(2)}} \leq \\
&& (N \delta t)^2 \lambda^2 \norm{H_{SB}}^2 (c_0 \lambda \norm{H_{SB}}+c_1 \norm{H_B+H_X}), \nonumber
\end{eqnarray}
where $c_0$ and $c_1$ are some small constants. For the second term we have (using Eqs.~(119) and (125) in \cite{ng_combining_2011}):
\begin{eqnarray}
&&\norm{\sum_{k\geq 3}{H_{\textrm{eff}}^{(k)}}} \leq \\
&&(N \delta t)^3 \lambda \norm{H_{SB}} (\lambda \norm{H_{SB}}+\norm{H_B+H_X})^3  \nonumber \\
&& \times (c_2 +c_3 (\lambda \norm{H_{SB}}+\norm{H_B+H_X}) N_{\rm DD} \delta t)), \nonumber
\end{eqnarray}
where $c_2$ and $c_3$ are some small constants. Denoting this upper bound by $\eta'$ (the right hand side of Eq.~\eqref{Eq:UpperDD}), the upper bound on the effective error strength of the combined DD and simulation procedure becomes $N_{\rm sim} \eta'$ plus the error we get from the simulation. The strongest of these errors is at order $\delta t^4$, and a bound on these terms is $c(N_{\rm sim} N_{\rm DD} \delta t)^4 \norm{H_X}$. (This comes from counting the number of terms in the Magnus expansion and considering the locality of $H_X$, so rather than having $\norm{H_X}^4$, we have a constant times $\norm{H_X}$.) Similarly, higher order error terms of the simulation can also be included up to any desired accuracy.

The strength of the Hamiltonian itself, after the pulses, becomes $N_{\rm DD} H_{X}$, and so after the simulation the desired Hamiltonian shows up at order $(N_{\rm DD} \delta t)^3$. Increasing $N_{\rm DD}$ makes the simulated Hamiltonian stronger, but the error associated with DD can become worse, as can be seen from the bound.

\subsection{Locality and strength of noise}

Local noise on the original Hamiltonian can act as effective nonlocal noise on the simulated Hamiltonian. This is true for the DD process, and also for the simulation process. Here, starting from the Hamiltonian in Eq.~\eqref{eq:totalHamiltonian}, we investigate how the locality and strength of the effective noise changes after applying pulses.
For generality, we assume that the system Hamiltonian is $k$-local and the system bath interaction, $H_{SB}$, consists of $l$-local terms on the system (for our construction $k=2,l=1$).

Using the Magnus expansion it is evident that the terms in  $\delta t ^m$ are in the form of $m-1$ nested commutators (basically, products of $m$ Hamiltonians).

\subsubsection{Growth of nonlocality of noise in higher order terms}
First, let us assume that the bath part of the interaction Hamiltonian  can be arbitrarily nonlocal: $\forall i\neq j: \, [B_\alpha^i , B_\beta^j]\neq 0$. Conjugating a segment of the evolution with  $u_a$ pulses, is equivalent to having a evolution with an effective Hamiltonian $H_i=H_{a,X} + H_B +\lambda  H_{a,SB}$, where:
\begin{eqnarray}
u_a e^{-i H \delta t} u_a^{\dagger}=e^{-i (H_{a,X} + H_B +\lambda  H_{a,SB})}.
\end{eqnarray}

 The pulses do not change the locality of the system Hamiltonian or the interaction Hamiltonian (i.e., the locality of $H_{a,X}$ and $H_{a,SB}$ is the same as that $H_X$ and $H_{SB}$). The effective Hamiltonian at order $\delta t^m$ would be the result of $m-1$ nested commutators of these $H_a$ segments. At this order in $\delta t$,  for any $q$, the terms of order $\lambda^q$ are at most $[ql+(m-q)(k-1)]$-local. The most nonlocal terms are $ml$-local, and they appear at order $\lambda^m$ (which is quite weak). We can have a $[m(k-1)+1]$-local term at order $\lambda^0$, which we use for the simulation.
 
 In the case of DD, if we choose the $\{u_a\}$ pulses to commute with the system Hamiltonian we get $H_a=H_{X} + H_B +\lambda H_{a,SB}$. In this case, all the $H_B$ and $H_X$ terms commute with each other and the $\lambda^0$ term vanishes. 

\subsubsection{The case of our construction}

For the case where the system Hamiltonian is $2$-local and the interaction Hamiltonian is 1-local ($k=2,l=1$), errors of order $\lambda^q$ for $\forall q\geq 1$ are $[ql+(m-q)(k-1)]=m$-local, and at order $\lambda^0$ the simulated Hamiltonian is $(m+1)$-local. So at order $\delta t^3$ ($m=3$) we have $3$-local noise and the 4-body simulated Hamiltonian. 
(See Appendix~\ref{app:estimNum} for an estimate of the number of error terms.)

In our construction, we use $N_{\rm sim}N_{\rm DD}$ pulses to simulate $h H_p$ at order $\delta t^3$ with errors also at order $\delta t^3$, with the errors of leading order $\lambda^2$. (These terms are at most $3$-local, and there are at most  $N_{\rm sim} \times N_{\rm DD}^3 \times (3N)^2$ such terms.) The leading order terms in the simulation error are of order $\delta t^4$ and of order $\lambda^0$. (These terms are at most $5$-local, and there are $O(N_{\rm sim}^4)$ such terms.)

\subsubsection{Geometrically local bath}

It is reasonable to assume that the dynamics of the bath and the way it interacts with the system can be characterized by some local Hamiltonians.  Assuming locality for the bath part of the interaction Hamiltonian and also the bath Hamiltonian, one can show that the stronger noise terms (lower powers of $\lambda$) are more geometrically local. For example this means that the $3$-local errors showing up in $\delta t^3$ are more concentrated in a geometrically local region of the lattice, rather than showing up in any three locations. The surface code Hamiltonian (and topological codes in general) are expected to perform better if the errors are more geometrically correlated.

Now we assume locality for the bath part of the interaction Hamiltonian. We assume that the terms $B_i^\alpha$ and $B_j^\beta$ from the interaction Hamiltonian commute with each other if the corresponding system qubits are at least distance $r$ from each other.  
\begin{eqnarray}
\forall i,j \quad d(i,j) \geq r \Rightarrow \, [B_\alpha^i , B_\beta^j]= 0.
\end{eqnarray}

Again, looking at the nested commutators (we will add the effect of the bath Hamiltonian later), we see that errors of order $\lambda$ consist of a single term from $H_{SB}$ commuted with one or more terms from the system Hamiltonian. This error term therefore can be $[l+(m-1)(k-1)]$-local, but all the qubits it acts on are neighbors. In general, $[ql+(m-q)(k-1)]$-local errors in $\lambda^{q}$ can appear with the qubits separated by a distance of at most $(q-1)r$, and the geometrically local terms coming from the system Hamiltonian can only grow from this base of qubits.

We need to also include the effects of the bath Hamiltonian.  $H_B$ does not itself increase the non-locality of the interaction Hamiltonian terms, but it can connect terms in different local regions and so affect the spread of error. One can separately assume geometric locality for the bath Hamiltonian and repeat the argument above to bound the spread of errors. Namely, we may assume that $H_B=\sum{h_B}$, where each term can have a nontrivial effect on the bath part of $H_{SB}$ corresponding to qubits separated by a distance of at most $r'$:
\begin{eqnarray}
\forall h_B, \, d(i,j) &\geq& r': \, \nonumber \\ 
&\Rightarrow & [B_\alpha^i, h_B]=0 \textnormal{ or } [B_\alpha^j, h_B]=0.
\end{eqnarray}
If there are $b$ copies of $H_B$ in the $m$ nested commutators, at order $\lambda^d$ this results results in at most $[d l+(m-d-b)ˆ—(k-1)]$-local term that can spread up to a distance of $b r' + (d-2)r$ with $d\geq 2$.

\subsection{DD pulses for a local bath }
Suppose we want to protect $N$ qubits on a line from any possible multiqubit error. The usual twirling lemma would suggest applying all $4^N$ possible pulses. One can ask: what if each error term is supported on at most $l$ neighboring qubits?  Here we discuss another symmetrization lemma to lower the number of pulses to just $4^l$ multiqubit pulses, independent of $N$ (benefiting from the parallelism in applying pulses).

The construction of the pulses is as follows: list all $4^l$ possible Pauli operators on the first $l$ qubits. Now extend each pulse by periodically repeating the Pauli operator on each successive set of $l$ qubits such that  each of the pulses has the same Pauli on qubits $i$ and $i+l$.

A special case is 1-local noise on $N$ qubits ($l=1$, so the period is one). As we saw we can average out all the errors by considering just four pulses: $I^{\otimes N},X^{\otimes N},Y^{\otimes N},Z^{\otimes N}$.

Clearly we can similarly generalize this construction to higher dimensional lattices:

\begin{mylemma} \emph{(Symmetrization lemma: Local noise)}
	Assume that the support of each error term on a $D$-dimensional regular lattice is contained in a hypercube of size $l^D$ qubits. Then with $4^{l^D}$ pulses, independent of the size of the lattice, all the error terms can be averaged out.
\end{mylemma}

Applying this result to our 2-dimensional square lattice, we only need to construct the sequences by considering all possible Pauli operators on the qubits in squares of length  $l$ (there are $4^{l^2}$ corresponding pulses), and then cover the lattice with parallel use of these patterns.


For example, for $r=1$ (bath operators from the interaction Hamiltonian, $B_i^\alpha$ and $B_j^\beta$, commute unless qubits $i$ and $j$ are nearest neighbors) all the errors of order $\delta t^3$ are strings of 3 Pauli operators. We can define a square of $3\times 3$ qubits, and apply all Pauli operators that commute with all the $XX$ interactions in that square ($8$ independent interaction of such type). At most $4^{9}/2^8$ steps are needed (all the possible Pauli's operators on $9$ qubits that commute with $8$ independent Pauli terms).

Therefore, assuming locality for the bath, all error terms of order $\delta t^3$ can be removed with a number of pulses independent of the size of the grid. Of course this is a worst case analysis (assuming that any kind of three-Pauli error can be generated), but when needed one can work out the details of the generated errors and find the pulses to remove them, resulting in much shorter sequences.


\section{Error suppression} \label{sec:errorsup}

Now we analyze the effect of the simulated energy gap on the noise. For simplicity, rather than adding the simulation of the desired computation to this picture, we just repeat the process of simulating $h H_p \delta t^3$ a total of $k$ times (a quantum memory). Each repetition of the simulation represents a time step $\Delta t=\delta t^3$. We recall that the combination of DD and simulation described earlier generates the following effective Hamiltonian in one cycle:
\begin{eqnarray}
e^{-i [h H_p+H_B/\delta t^2-g V(\delta t)] \Delta t},
\end{eqnarray}
where $h$  is the strength of the simulated Hamiltonian and $V(\delta t)=\sum_{a=0}^{\infty} {V_a \delta t^a}$ is the effective error Hamiltonian. Each term $V_a$ can be decomposed according to the location of the system part: $V_a=\sum_{i,a}{V_i^a}$, and $g$ is the overall scale [normalizing $V(\delta t)$]. The ratio between $h$ and $g$ quantifies the energy penalty as we will see later.

We can think of the effective Hamiltonian as a constant Hamiltonian $h H_p+H_B/\delta t^2-g V(\delta t)$, that is on for time $k \Delta t$. The goal is to see how much error suppression we get from this process, assuming that we start in the ground subspace. To do so we bound the quantity:
\begin{eqnarray}
\norm{U(k \Delta t) P-  U_{\rm ideal}(\Delta t)P},
\end{eqnarray}
where $P$ is the projector onto the codespace, and $U(t)$ is the evolution generated by $H(t)={h H_P(t)}+H_B/\delta t^2+ V(t)$. This difference bounds how much the noisy evolution can make the state deviate from the ideal evolution represented by $U_{\rm ideal}(t)\equiv e^{-i (h H_p+H_B/\delta t^2) k t}$.

To bound this difference, we start by moving to the interaction picture defined by the base Hamiltonian ${h H_P}+H_B/\delta t^2$. We denote the evolution corresponding to this base Hamiltonian by $U_P(t)$. Moving to the interaction picture with respect to the base Hamiltonian we have:
\begin{eqnarray}
	V_I(t)&=&U^{\dagger}_P(t) V(t) U_P(t)\ ,\\
	U_I(t)&=& U^{\dagger}_P(t) U(t)\ ,
\end{eqnarray}
where 
\begin{eqnarray}
i \dot{U}_I(t)=V_I(t)U_I(t)\ .
\end{eqnarray}
Integrating this we have 
\begin{eqnarray}
U_I^{\dagger}(k \Delta t)=I +i \int_0^{k \Delta t} {U_I^{\dagger}(t) V_I(t) dt}\ .
\end{eqnarray}

We note that 
\begin{eqnarray}
&&\norm{U(k \Delta t)P-U_P(k \Delta t)P} \hspace*{\fill} \nonumber\\
&&\hspace{35pt} \hfill =\norm{(I-U_I^{\dagger}(k \Delta t) )P}\nonumber \\
&&\hspace{35pt} =\norm{\int_0^{k \Delta t} \frac{d U_I^{\dagger}(t)}{dt} dt P}\nonumber \\
&&\hspace{35pt} =\norm{\int_0^{k \Delta t} U^{\dagger}_I(t) V_I(t) dt P}\ .
\label{eq:normdiff}
\end{eqnarray}
Following \cite{bookatz_error_2014}, we can define: 
\begin{eqnarray}
F(t)=\int_0^t V_I(\tau) d\tau P\ .
\end{eqnarray}
Integrating by parts, Eq.~\eqref{eq:normdiff} becomes:
\begin{eqnarray}
&&\norm{U(k \Delta t) P- U_P(k \Delta t)P} \nonumber \\
&=&\norm{U_I^{\dagger}({k \Delta t})F({k \Delta t})- i \int_0^{k \Delta t}{U_I^{\dagger}(\tau) V_I(\tau) F(\tau)d\tau}}\nonumber\\
&\leq& \norm{F({k \Delta t})}+\norm{\int_0^{k \Delta t}{U^{\dagger}(\tau) V(\tau) U_P(\tau) F(\tau)d\tau}}\nonumber\\
&\leq& \norm{F({k \Delta t})}+\int_0^{k \Delta t}{ \norm{V(\tau)}\norm{F(\tau)}d\tau}\ , \label{eq:Difference}
\end{eqnarray} 
where we used the triangle inequality and the unitary invariance of the operator norm.

$F(t)$ quantifies the averaging out of the interaction Hamiltonian by the rotations induced by the penalty Hamiltonian (in the interaction picture), and the reduction of this term implies that the total evolution becomes closer to the ideal evolution (the difference between the actual and ideal evolutions goes to zero if $\norm{F}$ goes to zero \cite{bookatz_error_2014,PhysRevLett.118.030504}).

To bound this term, we can evaluate:
\begin{eqnarray}
F(t')= g&&\int_0^{t'}U^\dagger_P(\tau)
\sum_{i,a}{V_i^a \delta t^a}  U_P(\tau) P d\tau\ .
\end{eqnarray}

We denote by $c_i^a$ the number of stabilizer generators in the simulated Hamiltonian $H_p$ that anticommute with the error $V_i^a$. 
Using this, and also the fact that {$H_p P=\epsilon_0P$}, we have:
\begin{eqnarray}
F(t')=g \int_0^{t'}&&e^{+i (H_B/\delta t^2) \tau}\\
&&\times \sum_{i,a}{V_i^a \delta t^a}e^{-i (H_B/\delta t^2) \tau}   e^{-i 2 h \epsilon_0 c_i^a \tau} P d\tau\ . \nonumber
\end{eqnarray}

Breaking  this sum into two parts depending on whether $c_i^j$ is zero or nonzero, and then integrating by parts, we get:
\begin{eqnarray}
&&\norm{F(k\Delta t)} \leq \nonumber \\
&&\frac{g}{h } \sum_{\substack{i,a \\ c_i^a \neq 0}}{\frac{1}{c_i^a (2 \epsilon_0)} (2 \norm{V_i^a \delta t^a}+k \Delta t \norm{[V_i^a, H_B] \delta t^{a-2}})} \nonumber\\
&&+g k \Delta t  ||\sum_{\substack{i,a \\ c_i^a = 0}}{V_i^a \delta t^a}||\ .
\end{eqnarray}

It is important to note that the first term, with the detectable errors (with $c_i^a \geq 1$), is suppressed by a factor of {$g/h$}. So for large values of $h$, we get error suppression, and the total evolution becomes closer to the one with no error. {This error suppression is similar to the error suppression we get from an ideal implementation of $H_p$ (rather than its simulation). This becomes clearer when we perform a similar calculation for the Hamiltonian $H_p$ with energy penalty $E_p$, in the presence of a local system-bath interaction Hamiltonian $\lambda H_{\rm SB}$ for a duration of $T_0$. Similar to the simulation case, using Eq.~\eqref{eq:Difference}, we can bound
\begin{eqnarray}
\norm{e^{-i (E_p H_p+H_B+\lambda H_{SB} )T_0 }P- e^{-i (E_p H_p+H_B)T_0 } P} \nonumber \\
\leq \norm{F_{\rm ideal}(T_0)}+\int_0^{T_0}{ \norm{\lambda H_{SB}}\norm{F_{\rm ideal}(\tau)}d\tau}\ ,  
\label{ideal}
\end{eqnarray}
where we have defined $F_{\rm ideal}(t)$ as:
\begin{eqnarray}
F_{\rm ideal}(t)= \int_0^{t}&&e^{+i (E_p H_p+H_B) \tau} \nonumber \\
&&\times (\lambda H_{SB})  e^{-i (E_p H_p+H_B) \tau} P d\tau\ .
\end{eqnarray}}

{Decomposing the interaction Hamiltonian according to the location of the system part, $H_{SB}=\sum_{i}{ h_{\rm SB}^i}$, we get: 
\begin{eqnarray}
&&\norm{F_{\rm ideal}(t)} \leq \nonumber \\
&&\frac{\lambda}{E_p } \sum_i{\frac{1}{e^i (2\epsilon_0)} (2 \norm{h_{\rm SB}^i}+t \norm{[h_{\rm SB}^i, H_B]})}\ ,
\end{eqnarray}
where $e^i$ denotes the number of stabilizer generators in the ideal $H_p$ that anticommute with the error term $h_{\rm SB}^i$. In this case, by construction, we always have $e^i \neq 0$.}

{This shows that the bound in Eq.~\eqref{ideal} actually contains a suppression factor of $\lambda/E_p$. Clearly, in the simulated case, $g/h$ plays the role of $\lambda/E_p$.}

{In contrast to the ideal case where $\forall i: e^i\neq 0$, in the simulation not all the terms of the effective error are suppressed by the effective Hamiltonian.} For the chosen $H_p$, the only errors that commute with all the stabilizers (and so have $c_i^a=0$) are (I) loops of $X$ around $A_v$ operators and loops of $Z$ around $B_p$ operators; and (II) logical operators, i.e., chains connecting boundaries. These errors are not suppressed by this mechanism, as can be seen from the second term in the bound.

Fortunately, the type I errors are not (too) destructive, {as they are just product of the stabilizers of the code. So, these terms in the effective error do not cause logical errors. The only effect they can have is changing the strength of each simulated stabilizer slightly} (changing $h$ to $h\pm \delta h$). The first error of this type is $4$-local and so occurs at order $O(\delta t^4)$. Comparing to the Hamiltonian itself, which occurs at order $O(\delta t^3)$, it is at least a factor of order $\delta t$  weaker. Also, the effect of these errors is expected to average out, as the sign of the errors changes, and so the effective $\delta h$ should be small.

The situation is different for the logical errors. If the distance of the code is $d$, these errors happen in $O(\delta t^d)$ or higher, which is small for large $d$. While this error is small, to have an arbitrary long computation it is necessary to correct possible errors before they accumulate into logical errors. This is done by active error correction.

\section{Summary and Conclusions} 
\label{sec:conclusion}

High-weight Hamiltonians are frequently used in designing quantum algorithms, especially when the goal is to provide protection against noise. Implementing such high-weight Hamiltonians is experimentally challenging. One approach is to simulate such interactions using resources that are easier to implement. However, this simulation procedure itself can spread the noise and convert it to some effective geometrically correlated noise on the simulated system.

In this work we proposed combining techniques from dynamical decoupling and quantum simulation to simulate high-weight Hamiltonians such that the simulated Hamiltonians are stronger than the new effective noise. The ratio of the strength between the simulated Hamiltonian and the effective noise depends on the strength of the original noise on the resources used and also on the specific type of DD and simulation techniques applied. The spread of the effective noise on the simulated system depends on the locality of the bath and the system-bath interaction Hamiltonians. The reasonable assumption that these Hamiltonians are geometrically local, guarantees that the dominant terms in the effective noise Hamiltonian spread in a geometrically local region. Topological codes are expected to perform well in the presence of these types of geometrically correlated errors. For this reason we chose the surface code Hamiltonian and showed how to simulate the time dependent deformation of this Hamiltonian to perform universal quantum computation. Our analysis provided the details specific to this Hamiltonian, but the method is general. Similar to any other scheme performing universal quantum computation on a 2D grid, active error correction is necessary to guarantee fault tolerance. But as we showed explicitly, an energy gap is maintained during the simulation against the strongest errors generated in the process. The presence of this energy gap reduces the number of cycles of active error correction necessary during the simulation of the computation.

We expect that the methods proposed in this work can also be used to simulate the Hamiltonians that use subsystem codes to reduce the non-locality, either for surface code Hamiltonians \cite{Bravyi:2013:SS} or general Hamiltonian-based quantum computation \cite{PhysRevLett.118.030504}.

An interesting problem for future work is to consider other types of resources and simulation methods. One example is to  consider simulation using perturbative gadgets and investigating methods to reduce the effect that local noise on a Hamiltonian has on the effective Hamiltonian in the low energy spectrum.

In all the constructions in this work, we only used the simplest form of DD and simulation techniques. One can expect to gain performance improvements by using more complex DD pulses and simulation techniques. It then becomes more important to consider the effect of the imperfection and noise on the pulse sequences and their timing. 

More generally, it is interesting to design methods that are natively optimized to generate the largest ratio between the strength of the simulated system Hamiltonian and the strength of the effective noise.


\acknowledgments
The authors acknowledge useful conversations with Yi-Cong Zheng. This work was supported under ARO MURI Grant Nos. W911NF-11-1-0268 and W911NF-15-1-0582; by NSF Grant No. CCF-1421078; and by an IBM Einstein Fellowship at the Institute for Advanced Study.

\appendix

\section{Alternative Connectivity} \label{App:alternativeconnectivity}

An alternative connectivity for the Hamiltonian on the square grid is when there is only nearest-neighbor coupling.
In this case the Hamiltonian would become:
\begin{eqnarray}
H_X=\sum_{\langle i,j\rangle}{X_i X_j}\ .
\end{eqnarray}

\begin{figure}[!t]
\begin{center}
\includegraphics[height=0.25\textheight]{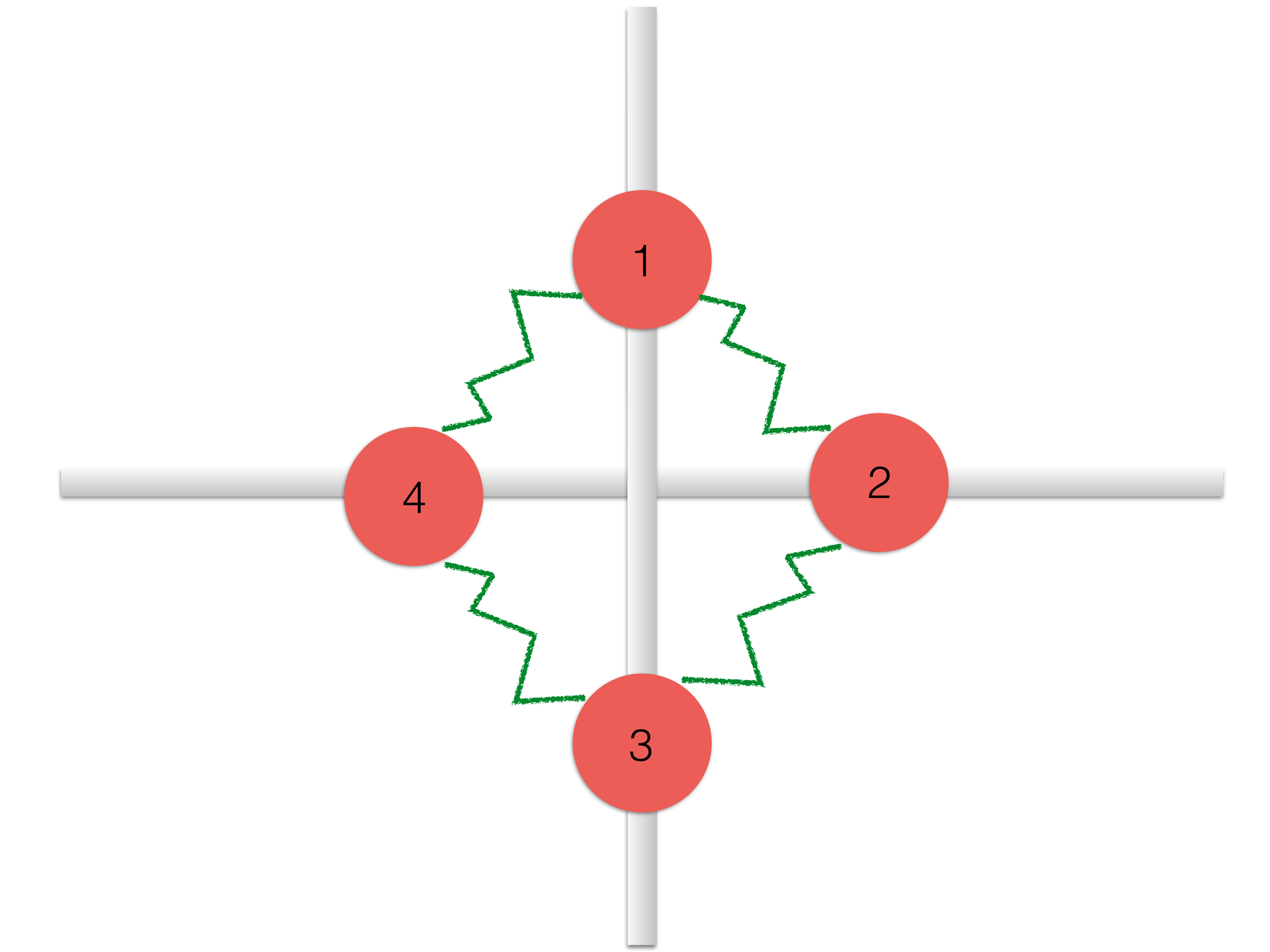}
\caption{Labeling of qubits connected to a vertex}
\label{Fig:OneVertix_Onlynearest}
\end{center}
\end{figure}

Here we show how to use the Hamiltonian $H_x= (X_1+X_3)(X_2+X_4)$ to simulate the Hamiltonian $X_1 X_2 X_3 X_4$ on qubits connected to a vertex, for an effective time $\Delta t$. (See Figure \ref{Fig:OneVertix_Onlynearest}.) We denote the Hadamard gate by $W$, and the Phase gate by $S$.

Now,
\begin{eqnarray}
(S_1 S_2) e^{-i H_X \delta t}(S_1 S_2) =e^{-i (Y_1+X_3)(Y_2+X_4) \delta t} 
\end{eqnarray}
generates $H_a=(Y_1+X_3)(Y_2+X_4)$. Likewise, 
\begin{eqnarray}
(W_1 S_2) e^{-i H_X \delta t}(W_1 S_2) =e^{-i (Z_1+X_3)(Y_2+X_4) \delta t}
\end{eqnarray}
generates $H_b=(Z_1+X_3)(Y_2+X_4)$, and
\begin{eqnarray}
(W_2) e^{-i H_X \delta t}(W_2) =e^{-i (X_1+X_3)(Z_2+X_4) \delta t}
\end{eqnarray}
generates $H_c=(X_1+X_3)(Z_2+X_4)$. 

All of these Hamiltonians can be negated using $X_2 Z_4$ pulses. Using the commutator method given in Eq.~\eqref{Eq:commutatorMethod}, we will have the the desired Hamiltonian $8 X_1 X_2 X_3 X_4$ plus some extra terms. Again, by doubling the number of pulses and conjugating the new ones by $Z_1 Y_2$, we can get rid of all the extra terms, and so the desired Hamiltonian becomes
\begin{eqnarray}
e^{-i 16 X_1X_2X_3X_4  \delta t^3} +O(\delta t^4). \nonumber
\end{eqnarray}

So using $N_{X}=20$ pulses, and applying the Hamiltonian $H_x$ 
for a time $N_{X} \delta t$ we can simulate $16 X_1X_2X_3X_4$ at order $\delta t^3$ with errors of order $\delta t^4$. 

Generating this Hamiltonian on a grid can be done using the methods from Sec.~\ref{sec:noisefree} with slight modifications.

\section{Magnus expansion for piecewise constant Hamiltonian} \label{app:Magnus}
A good reference introducing the Magnus expansion is \cite{ng_combining_2011}.
Here are a few low-order terms for piecewise constant Hamiltonians \cite{santos2008advantages}:
\begin{eqnarray}
e^{ -i H_n \delta t}\cdots e^{ -i H_2 \delta t}e^{ -i H_1 \delta t}=e^{ -i T_n H_{\textrm{eff}}(T_n)}.
\end{eqnarray}
Here $T_n= n \delta t$ and $H_{\textrm{eff}}= \sum_{k=0}^{\infty}{H^{(k)}_{\textrm{eff}}}$, with
\begin{eqnarray}
H^{(0)}_{\textrm{eff}}(T_n)&=& \frac{\delta t}{T_n} \sum_{k=1}^{n}{H_k},\\
H^{(1)}_{\textrm{eff}}(T_n)&=& -\frac{i (\delta t)^2}{2T_n} \sum_{l=2}^{n}\sum_{k=1}^{l-1}{[H_l,H_k]},\\
H^{(2)}_{\textrm{eff}}(T_n)&=&\\
-\frac{ (\delta t)^3}{6T_n}&&\left\{\sum_{m=3}^{n}\sum_{l=2}^{m-1}\sum_{k=1}^{l-1}{[H_m,[H_l,H_k]+[H_m,H_l],H_k]} \right. \nonumber \\
&+& \left.  \frac{1}{2} \sum_{l=2}^{n}\sum_{k=1}^{l-1}{[H_l,[H_l,H_k]+[H_l,H_k],H_k]]}\right\}. \nonumber
\end{eqnarray}

\section{Proof of lemma 1}
\label{app:Twirling}
\textbf{Lemma 1.} {(Symmetrization lemma: Protecting interaction)} \emph{
Let $P_1,\ldots,P_n$ be commuting Pauli operators, and let $\mathcal{P}$ be the group generated by these Pauli operators. Denote the set of all Pauli operators that commute with all the $P_i$s by $\mathcal{N}(\mathcal{P})$ {(the normalizer of $\mathcal{P}$)}. Then for any Pauli operator $s$,
\begin{eqnarray}
&&\forall s \notin \mathcal{P:} \sum_{g\in \mathcal{\mathcal{N}(\mathcal{P})}}{g s g^{\dagger}}=0,\\
&&\forall s \in \mathcal{P:} \sum_{g\in \mathcal{\mathcal{N}(\mathcal{P})}}{g s g^{\dagger}}=s |\mathcal{\mathcal{N}(\mathcal{P})}|.
\end{eqnarray} }

\begin{proof}

The $s \in \mathcal{P}$ case is trivial. Assume $s \notin \mathcal{P}$ and $s \notin \mathcal{N}(\mathcal{P})$, which means that $s$ anticommutes with at least one element of $\mathcal{P}$ and of $\mathcal{N}(\mathcal{P})$. Denoting a set of independent generators of $\mathcal{P}$ by $\{ P_i \}_{i=1}^{m}$, there exists a (non-identity) $P_*$ in this set such that $\{P_*,s\}= 0$. Obviously $\forall g \in \mathcal{N}(\mathcal{P}) \Rightarrow P_* g \in \mathcal{N}(\mathcal{P})$. From this we conclude that $g$ and $P_* g$ are two distinct elements of $\mathcal{N}(\mathcal{P})$ such that one commutes with $s$, and the other one anticommutes with it.\footnote{There are two cases depending on whether $s$ commutes or anticommutes with $g$. If $\{s,g\}=0,\{s,P_*\}=0$ then $P_*gs = -P_*sg = sP_*g$, i.e., $[s,P_*g]=0$. Likewise, $[s,g]=0,\{s,P_*\}=0 \Rightarrow \{s,P_*g\}=0$.} 
Going through all the elements of $\mathcal{N}(\mathcal{P})$ (and using $P_{*}^2=I$), half the elements commute with $s$ while the other half anticommute with $s$, and so the result holds in this case.

For the case that $s \notin \mathcal{P}$ and $s \in \mathcal{N}(\mathcal{P})$: elements of  $\mathcal{N}(\mathcal{P})$ can be generated using $\{ P_i \}$ and $n-m$ extra pairs of conjugate generators, where $n$ is the number of qubits. Let us call these pairs $(\hat{X}_j, \hat{Z}_j)$, where $1 \leq j \leq n-m$. Note that
$s \notin \mathcal{P}$ and $s \in \mathcal{N}(\mathcal{P})$ means that $s$ contains at least one element from these pairs, let us say an element from the $j_*$-th pair. Now we notice that $\forall g \in \mathcal{N}(\mathcal{P}) \Rightarrow \{g, g \hat{X}_{j_*},g \hat{Z}_{j_*},g \hat{X}_{j_*}\hat{Z}_{j_*}\}  \in \mathcal{N}(\mathcal{P})$. All these elements are distinct, and two of them commute with $s$ while two anticommute. Therefore, again the elements of $\mathcal{N}(\mathcal{P})$ can be partitioned into half commuting and half anticommuting, which completes the proof.
\end{proof}

\section{The number of pulses to average out multiqubit errors grows with the number of qubits} \label{app:DDMultiqubitGrows}

A simple example showing that the number of pulses to average out multi-qubit errors grows with the number of qubits is the following set of two-qubit errors:
\begin{eqnarray}
\{ Y_i Y_j : 1 \le i < j \le N \}
\end{eqnarray}
A simple lower bound on the number of pulses needed to average out all of these errors can be derived by noticing that each $Y_i$ has to have a distinct pattern of commutation/anticommutation with the pulses: if $Y_i$ and $Y_j$ have the same pattern, then all the pulses will commute with $Y_i Y_j$, so applying pulses will leave this term unchanged. From this we conclude that number of distinct patterns has to be greater than $N$, which means that the number of pulses needed to average out all the error terms has to increase with the number of qubits. If $|P|$ is the number of pulses, this means that $2^{|P|} \geq N$, and so ${|P|} \geq \log_2{N}$. For combinatorial approaches using $O(N)$ pulses to achieve first-order decoupling see Refs.~\cite{PhysRevA.65.042309,Wocjan:2006aa} and chapter 15 of Ref.~\cite{lidar2013quantum}. 



\section{Estimating the Number of Error Terms} 
\label{app:estimNum}
We provide a worst-case estimate of the number of error terms that occur at order $\delta t^m$ while having the simulated Hamiltonian appear at order $\delta t^{m+1}$.

Using our construction, we first apply  $N_{\rm DD}$ pulses to protect the system Hamiltonian and push the errors to order $\delta t^m$. At this order there will be roughly $N_{\rm DD}^m$ terms of $m-1$ nested commutators, with each term of the form $H_i=H_{X} + H_B + \lambda H_{i,SB}$. At order $\lambda^0$ there is no error, as all terms of $H_{X}$ and $H_B$ commute with each other. At order $\lambda^1$ there are $3 N$ terms ($N$ being the number of qubits, and $3N$ being the number of terms), multiplied by a constant depending on the connectivity degree of $H_X$, as the entangling Hamiltonian can only expand an existing string of errors locally to neighboring qubits. (The same argument is true for higher powers of $\lambda$.) Thus, at order $\lambda^1$ the total number of terms is of order $N^m_{\rm DD} \times 3N$. Notice the growth with the number of qubits. At order $\lambda^q$ there are $N_{\rm DD}^m (3N)^q$ terms. (Recall that we can remove all the errors of order $\lambda^1$  by doubling the number of pulses.)

Therefore, to simulate the desired nonlocal Hamiltonian at order $\delta t^{m'}$, we apply $N_{\rm sim}$ of these sequences of pulses consecutively. The number of error terms generated by the simulation process at order $\delta t^{m'+1}$ is $N^{m'}_{\rm sim}$, which as we saw earlier can be $[m'(k-1)+1]$-local at order $\lambda^0$. Also, the number of error terms resulting from DD will be multiplied by $N_{\rm sim}$.

\bibliographystyle{apsrev4-1}
\bibliography{Ref1}

\end{document}